\documentclass[twocolumn,showpacs,superscriptaddress,amsmath,amssymb,aps,pra]{revtex4-2}
\usepackage{graphicx}
\usepackage{CJKutf8}
\usepackage{dcolumn}
\usepackage{amsmath,bm}
\usepackage{epstopdf}
\usepackage[mathlines]{lineno}
\usepackage{color}
\usepackage[colorlinks=true,linkcolor=blue,citecolor=magenta,urlcolor=cyan]{hyperref}

\usepackage{tikz,xcolor,hyperref}

\definecolor{lime}{HTML}{A6CE39}
\DeclareRobustCommand{\orcidicon}{%
	\begin{tikzpicture}
	\draw[lime, fill=lime] (0,0) 
	circle [radius=0.16] 
	node[white] {{\fontfamily{qag}\selectfont \tiny ID}};
	\draw[white, fill=white] (-0.068,0.105) 
	circle [radius=0.007];
	\end{tikzpicture}
	\hspace{-2mm}
}

\foreach \x in {A, ..., Z}{%
	\expandafter\xdef\csname orcid\x\endcsname{\noexpand\href{https://orcid.org/\csname orcidauthor\x\endcsname}{\noexpand\orcidicon}}
}

\begin{document}
\title{Two-band description of the strong `spin'-orbit coupled one-dimensional hole gas in a cylindrical Ge nanowire}
\author{Rui\! Li~(\begin{CJK}{UTF8}{gbsn}李睿\end{CJK})\orcidA{}}
\email{ruili@ysu.edu.cn}
\affiliation{Key Laboratory for Microstructural Material Physics of Hebei Province, School of Science, Yanshan University, Qinhuangdao 066004, China}

\author{Xin-Yu\! Qi~(\begin{CJK}{UTF8}{gbsn}齐新雨\end{CJK})}
\affiliation{Key Laboratory for Microstructural Material Physics of Hebei Province, School of Science, Yanshan University, Qinhuangdao 066004, China}
\begin{abstract}

The low-energy effective Hamiltonian of the strong `spin'-orbit coupled one-dimensional hole gas in a cylindrical Ge nanowire in the presence of a strong magnetic field is studied both numerically and analytically. Basing on the Luttinger-Kohn Hamiltonian in the spherical approximation, we show this strong `spin'-orbit coupled one-dimensional hole gas can be accurately described by an effective two-band Hamiltonian $H^{\rm ef}=\hbar^{2}k^{2}_{z}/(2m^{*}_{h})+\alpha\sigma^{x}k_{z}+g^{*}_{h}\mu_{B}B\sigma^{z}/2$, as long as the magnetic field is purely longitudinal or purely transverse. The explicit magnetic field dependent expressions of the `spin'-orbit coupling $\alpha\equiv\alpha(B)$ and the effective $g$-factor $g^{*}_{h}\equiv\,g^{*}_{h}(B)$ are given. When the magnetic field is applied in an arbitrary direction,  the two-band Hamiltonian description is still a good approximation. 
\end{abstract}
\date{January 16, 2023}
\maketitle

\section{Introduction}

In the framework of the effective mass approximation, the band dispersions near the top of the valence band of semiconductors are well described by the Luttinger-Kohn Hamiltonian~\cite{PhysRev.97.869,PhysRev.102.1030}. The holes in the valence band have spin $3/2$, and in most circumstances a four-band Luttinger-Kohn Hamiltonian is enough to account for the various hole related phenomena in semiconductors~\cite{PhysRevB.31.888,PhysRevB.43.9649,PhysRevB.95.075305}. For semiconductors such as Ge, GaAs, InAs, and InSb, a spherical approximation can be further made to the Luttinger-Kohn Hamiltonian~\cite{PhysRevB.40.8500,PhysRevB.8.2697,Semina:2015aa,WU201061}. The spherical approximation greatly reduces the amount of calculation in the theoretical investigations, and especially some hole models of quantum dot~\cite{PhysRevB.42.3690}, quantum wire~\cite{PhysRevB.42.3690,sweeny1988hole}, and quantum well~\cite{PhysRevB.36.5887} are now exactly solvable.

Quasi-two-dimensional (2D) hole gas achieved in semiconductor quantum well or heterostructure has been studied for decades~\cite{PhysRevB.31.888,ando1985hole,RASHBA1988175,PhysRevB.52.11132,Hendrickx:2020ab,Hendrickx:2020aa,Scappucci:2021vk}. The properties of the 2D hole gas are relatively well understood. The lowest subband dispersion is approximately parabolic, and the subband minimum is at the center of the momentum space~\cite{winkler2003spin,PhysRevB.36.5887}. In the presence of an electrical field perpendicular to the quantum well, the Rashba spin-orbit coupling is found to be third order in momentum~\cite{winkler2003spin,PhysRevB.62.4245}. Note that recently a linear in momentum Rashba spin-orbit coupling is shown to exist in some low symmetry quantum wells~\cite{PhysRevB.103.085309,PhysRevB.105.L161301}. Hole spins localized in quasi-2D quantum dot have potential applications in quantum computing. The electrical spin manipulation~\cite{PhysRevLett.98.097202,Wang:2021wc,PhysRevB.103.125201}, the phonon induced spin relaxation~\cite{PhysRevLett.95.076805}, and the spin-cavity coupling~\cite{PhysRevB.102.205412} have been extensively studied.

Quasi-1D hole gas achieved in semiconductor Ge nanowire is attracting increasing interest recently~\cite{PhysRevLett.101.186802,Watzinger:2018aa,PhysRevResearch.3.013081,PhysRevLett.112.216806,Froning:2021aa,Wang:2022tm,PhysRevB.88.241405,PhysRevB.99.115317,PhysRevB.104.235304,PhysRevB.105.075308,PhysRevB.106.235408}. The hole gas in 1D Ge nanowire exhibits peculiar properties that cannot be anticipated from the counterpart properties in 2D. The subband minimum of the 1D hole gas is not at the center of the momentum space~\cite{PhysRevB.84.195314,RL2021}. Instead, there are two symmetrical minimums approximately located at $|k_{z}R|\approx0.52$~\cite{RL2022a}. In the presence of an electric field perpendicular the nanowire, the Rashba spin-orbit coupling is found to be linear in momentum~\cite{PhysRevB.84.195314,PhysRevB.97.235422,PhysRevB.90.195421,PhysRevLett.119.126401}.  Recent experiments have also revealed some interesting hole properties in 1D. Very short spin-orbit lengths of 3-20 nm are reported~\cite{PhysRevLett.112.216806,Froning:2021aa,Wang:2022tm}, the effective longitudinal $g$ factor can be as large as the transverse $g$-factor~\cite{Froning:2021aa}, and the electrical spin manipulation frequency can be as large as several hundreds of MHz~\cite{Froning:2021aa,Wang:2022tm}.

The peculiar low-energy subband structure of the 1D hole gas in the absence of both electric and magnetic fields, i.e., two mutually displaced parabolic curves with an anticrossing at $k_{z}R=0$~\cite{PhysRevB.84.195314,RL2021}, indicates there is a natural strong hole `spin'-orbit coupling. Note that each dispersion curve is two-fold degenerate, i.e., spin degeneracy~\cite{PhysRevB.84.195314,PhysRevB.84.121303,RL2021}. We simply lift this spin degeneracy by using an strong magnetic field, such that a pure (without hole spin degeneracy) `spin'-orbit coupled 1D hole gas can be obtained~\cite{RL2022a,RL2022b}. In this paper, basing on the Luttinger-Kohn Hamiltonian in the spherical approximation, we show both numerically and analytically this strong `spin'-orbit coupled 1D hole gas is accurately described by an effective two-band Hamiltonian $H^{\rm ef}=\hbar^{2}k^{2}_{z}/(2m^{*}_{h})+\alpha\sigma^{x}k_{z}+g^{*}_{h}\mu_{B}B\sigma^{z}/2$, as long as the external magnetic field is purely longitudinal or purely transverse. We draw this conclusion from the fact that we can exactly divide the subband dispersions of the hole gas into two subsets, and each subset can be labeled with a definite hole spin index. When the external magnetic field is applied in a general direction, in most cases, the effective two-band Hamiltonian is still a good approximation. Note that when the real spin (for both electron and hole) is concerned, the Hamiltonian $H^{\rm ef}$ has been broadly used in the studies of the spin-orbit qubits~\cite{trif2008spin,RL2013,PhysRevB.88.241405} and the searching of Majorana fermions~\cite{PhysRevLett.105.077001,PhysRevLett.105.177002,PhysRevB.90.195421}. Therefore, we expect the hole `spin' explored here will have similar potential applications.

\section{Strong `spin'-orbit coupled 1D hole gas} 

Quasi-1D hole gas is experimentally achievable in a Ge/Si core/shell nanowire~\cite{Lu10046} or Ge hut wire~\cite{Gao2020AM}. Quasi-1D means the holes can move freely in one dimension, while are strongly confined in the other two dimensions. Here, for the convenience of theoretical calculations, we focus on the 1D hole gas achieved in an unstrained cylindrical Ge nanowire. We show the hole gas is naturally strong `spin'-orbit coupled when a strong magnetic field is used to lift the real spin degeneracy. The magnetic field can be applied either longitudinally or transversely with respect to the nanowire axis. The subband quantization of the hole gas is described by the Hamiltonian
\begin{eqnarray}
H&=&\frac{1}{2m}\left[\left(\gamma_{1}+\frac{5}{2}\gamma_{s}\right)\textbf{p}'^{2}-2\gamma_{s}(\textbf{p}'\cdot\textbf{J})^{2}\right]\nonumber\\
&&+2\kappa\mu_{B}{\bf B}\cdot{\bf J}+V(r),\label{eq_model}
\end{eqnarray}
where $m$ is the bare electron mass, $\gamma_{1}=13.35$, $\gamma_{s}=5.11$, and $\kappa=3.41$ are Luttinger parameters for Ge~\cite{PhysRevB.4.3460}, ${\bf p'}=-i\hbar\nabla+e{\bf A}$ is the momentum operator in the presence of a magnetic field, with ${\bf A}$ being the vector potential ${\bf B}=\nabla\times{\bf A}$, ${\bf J}=(J_{x}, J_{y}, J_{z})$ is a spin-$3/2$ vector operator, and $V(r)$ is the transverse ($xy$ plane) confining potential
\begin{equation}
V(r)=\left\{\begin{array}{cc}0,~&~r<R,\\
\infty,~&~r>R,\end{array}\right.\label{Eq_potential}
\end{equation}
with $R$ being the radius of the cylindrical Ge nanowire. Note that in our following calculations we have set $R=10$ nm, a typical and experimentally achievable nanowire radius~\cite{PhysRevLett.101.186802,PhysRevLett.112.216806}.

\begin{figure}
\includegraphics{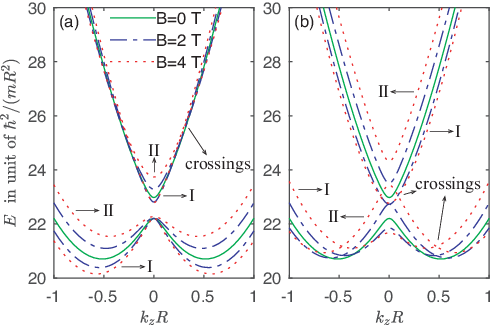}
\caption{\label{fig_subbands}Subband dispersions of the 1D hole gas in a strong longitudinal field (a) and transverse field (b). There are subband crossings at a fixed magnetic field (see the sites marked by the arrows).  Subband crossings imply the subband dispersions can be exactly classified into different spin subsets, e.g., one subset of suband dispersions belongs to spin up and the other subset belongs to spin down.}
\end{figure}

In the absence of the magnetic field (${\bf B}=0$), the hole Hamiltonian (\ref{eq_model}) is reduced to~\cite{PhysRevB.42.3690,PhysRevB.84.195314,PhysRevB.79.155323}
\begin{equation}
H_{0}=\frac{1}{2m}\left[\left(\gamma_{1}+\frac{5}{2}\gamma_{s}\right)\textbf{p}^{2}-2\gamma_{s}(\textbf{p}\cdot\textbf{J})^{2}\right]+V(r).\label{eq_model0}
\end{equation} 
This Hamiltonian is exactly solvable with the help of both the total angular momentum conservation and the hard-wall boundary condition~\cite{sweeny1988hole,PhysRevB.42.3690,RL2021}. The 1D hole gas has a peculiar low energy subband structure, there is a natural strong `spin'-orbit coupling in the hole subband dispersions. The lowest two subband dispersions are just two mutually displaced parabolic curves with an anticrossing at $k_{z}R=0$ (see the solid line in Fig.~\ref{fig_subbands}). Note that in the absence of magnetic field each dispersion line is two-fold degenerate, i.e., spin degeneracy~\cite{PhysRevB.84.195314,PhysRevB.84.121303,RL2021}.

When a strong magnetic field is applied, we can solve the Hamiltonian $H=H_{0}+H_{1}$ using quasi-degenerate perturbation theory~\cite{RL2022a,RL2022b}, where the perturbation $H_{1}$ contains all the magnetic terms ($\propto\,B$, $B^{2}$). The strong magnetic field lifts the two-fold degeneracy in the hole subband dispersions. Typical results of the subband dispersions in a strong longitudinal and transverse fields are shown in Figs.~\ref{fig_subbands}(a) and (b), respectively. At a fixed magnetic field, either longitudinal or transverse, there exist subband crossings [see Figs.~\ref{fig_subbands}(a) and (b)]. These subband crossings imply we can label the dispersion line with a definite spin index, i.e., we can classify the subband dispersions into different spin subsets. Each subset (lines I or II in Fig.~\ref{fig_subbands}) of the subband dispersions can be modeled by~\cite{RL2022a,RL2022b}
\begin{equation}
E(k_{z})=\frac{\hbar^{2}k^{2}_{z}}{2m^{*}_{h}}+\alpha\sigma^{x}k_{z}+\frac{g^{*}_{h}\mu_{B}B}{2}\sigma^{z}+const.,\label{eq_dispersion}
\end{equation}
where $\sigma^{x,z}$ are Pauli `spin' matrices. We also emphasize that the effective hole mass $m^{*}_{h}\equiv{}m^{*}_{h}(B)$, the Rashba~\cite{bychkov1984oscillatory} type `spin'-orbit coupling $\alpha\equiv\alpha(B)$, and the effective $g$-factor $g^{*}_{h}\equiv{}g^{*}_{h}(B)$ are all magnetic field dependent~\cite{RL2022a,RL2022b}. In the following, we show the correctness of Eq.~(\ref{eq_dispersion}) from the viewpoint of the effective Hamiltonian theory.

\section{Effective two-band Hamiltonian}

Following Ref.~\cite{PhysRevB.84.195314}, we can obtain an effective Hamiltonian describing the low-energy subband dispersions of the hole gas. The idea lies that we can solve both the eigenvalues and the eigenstates of Hamiltonian $H_{0}$ at the site $k_{z}=0$, then we treat the $k_{z}$ terms in Hamiltonian $H_{0}$ as a perturbation. Eventually, we would obtain an effective Hamiltonian valid at small $k_{z}R$ ($k_{z}R\ll1$). In the Hilbert subspace spanned by the lowest four eigenstates of $H_{0}(k_{z}=0)$, i.e., $|e+\rangle$, $|e-\rangle$, $|g+\rangle$, and $|g-\rangle$~\cite{PhysRevB.84.195314,RL2021}, one obtains~\cite{PhysRevB.84.195314}
\begin{eqnarray}
H^{\rm ef}_{0}&=&\frac{\hbar^{2}k^{2}_{z}}{4}\left(\frac{1}{m^{*}_{e}}+\frac{1}{m^{*}_{g}}\right)+C\frac{\hbar^{2}}{mR}k_{z}\tau^{x}s^{x}\nonumber\\
&&+\left[\frac{\hbar^{2}k^{2}_{z}}{4}\left(\frac{1}{m^{*}_{e}}-\frac{1}{m^{*}_{g}}\right)+\frac{\Delta}{2}\frac{\hbar^{2}}{mR^{2}}\right]\tau^{z}.\label{eq_Hamiltoniankz}
\end{eqnarray}
where ${\boldsymbol \tau}$ and ${\bf s}$ are Pauli matrices defined in the $\{g,e\}$ and $\{+,-\}$ subspaces, respectively, $\Delta\hbar^{2}/(mR^{2})$ is the level spacing between $|e\pm\rangle$ and $|g\pm\rangle$. The realistic values of  the parameters $m^{*}_{e,g}$, $C$, and $\Delta$ are given in Tab.~\ref{tab1}.

\begin{table*}
  \centering 
  \caption{The parameters of the effective Hamiltonian~\cite{PhysRevB.84.195314}. The values are calculated using the eigenstates of $H_{0}$ at $k_{z}R=0$, and those values given in the parentheses are obtained by a band fitting.}\label{tab1}
  \begin{ruledtabular}
  \begin{tabular}{cccccccccccc}
$m^{*}_{e}/m$\footnote{$m$ is the free electron mass}& $m^{*}_{g}/m$& $C$& $\Delta$& $Z_{1}$& $Z_{2}$& $Z_{3}$& $Z_{4}$& $X_{1}$& $X_{2}$& $X_{3}$& $X_{4}$\\
$0.054$ ($0.074$)& $0.043$~($0.074$)& $7.27$~($7.12$)& $0.77$& $0.75$& $0.82$& $-2.38$~($-3.62$)& $0.65$& $2.73$& $-0.18$& $8.04$~($4.66$)& $0.57$
\end{tabular}
\end{ruledtabular}
\end{table*}
\begin{figure}
\includegraphics{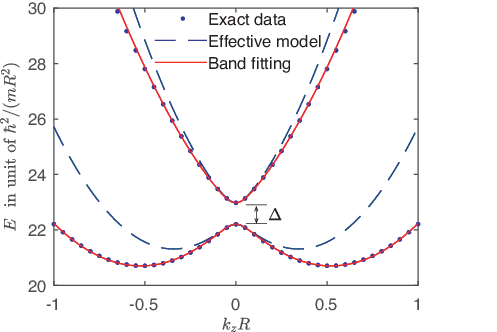}
\caption{\label{fig_B_0T}Comparison between the subband dispersions given by the exact method and that given by the effective Hamiltonian $H^{\rm ef}_{0}$ (\ref{eq_Hamiltoniankz}). A fitting to the exact subband dispersions gives rise to $m^{*}_{e}\approx\,m^{*}_{g}\approx0.074m$ and $C\approx7.12$.}
\end{figure}

A comparison between the exact subband dispersions and the subband dispersions given by the effective Hamiltonian (\ref{eq_Hamiltoniankz}) is shown in Fig.~\ref{fig_B_0T}. The effective Hamiltonian (\ref{eq_Hamiltoniankz}) indeed qualitatively reproduces the subband structure of the 1D hole gas. However, both the site and the magnitude of the subband minimum predicted by the effective Hamiltonian (\ref{eq_Hamiltoniankz}) are slightly different from the exact results. This discrepancy is due to the fact that $H^{\rm ef}_{0}$ is valid only at small $k_{z}R$ ($k_{z}R\ll1$), while the exact site of the band minimum is located at $|k_{z}R|\approx0.52$ (not very small)~\cite{RL2022a}. We also show the result of using Eq.~(\ref{eq_dispersion}) to fit the exact subband dispersions (see the solid line in Fig.~\ref{fig_B_0T}). The fitting gives rise to $m^{*}_{h}\approx\,m^{*}_{e}\approx\,m^{*}_{g}\approx0.074m$ and $C\approx7.12$. The band fitting well reproduces the low-energy subband structure of the 1D hole gas, such that in the following we have adopted the fitting parameters instead of the parameters calculated using the eigenstates at $k_{z}R=0$. Hence, we have
\begin{equation}
H^{\rm ef}_{0}\approx\frac{\hbar^{2}k^{2}_{z}}{2m^{*}_{h}}+C\frac{\hbar^{2}}{mR}k_{z}\tau^{x}s^{x}+\frac{\Delta}{2}\frac{\hbar^{2}}{mR^{2}}\tau^{z}.\label{eq_Hamiltoniankz2}
\end{equation}
Note that the parameter values given in the parentheses of Tab.~\ref{tab1} are obtained by a band fitting. We also emphasize that the reduction from Eq.~(\ref{eq_Hamiltoniankz}) to Eq. (\ref{eq_Hamiltoniankz2}) is an approximation. The kinetic term in the second line of Eq.~(\ref{eq_Hamiltoniankz}) is much smaller than the kinetic term in the first line, and hence is neglected.

\subsection{The strong longitudinal field case}

We now consider the effective Hamiltonian $H_{l}=H-H_{0}$ of the magnetic terms when a strong longitudinal (parallel to the nanowire axis) field is applied. We have
\begin{equation}
H^{\rm ef}_{l}=\mu_{B}B(Z_{1}s^{z}+Z_{2}\tau^{z}s^{z}-Z_{3}k_{z}R\tau^{y}s^{y})+\frac{e^{2}B^{2}R^{2}}{2m}Z_{4}\tau^{z},
\end{equation}
where $Z_{i}$ ($i=1,2,3,4$) are introduced to describe both the Zeeman and the orbital effects of the longitudinal field.  Note that the $Z_{1,2,3}$ terms have also been derived in Ref.~\cite{PhysRevB.84.195314}, the $Z_{4}$ term is induced by the orbital effects of the magnetic field proportional to $B^{2}$. The $Z_{4}$ term cannot be neglected because here we are considering strong magnetic fields. The realistic values of $Z_{1,2,3,4}$ are given in Tab.~\ref{tab1}. Interestingly, the $4\times4$ effective Hamiltonian $H^{\rm ef}=H^{\rm ef}_{0}+H^{\rm ef}_{l}$ is block diagonalized. In the Hilbert subspace spanned by $|e+\rangle$ and $|g-\rangle$, we have one effective two-band Hamiltonian
\begin{eqnarray}
H^{\rm ef}&=&\frac{\hbar^{2}k^{2}_{z}}{2m^{*}_{h}}+\left(C\frac{\hbar^{2}}{mR}+Z_{3}\mu_{B}BR\right)k_{z}\sigma^{x}\nonumber\\
&&+\left(\frac{\Delta}{2}\frac{\hbar^{2}}{mR^{2}}+Z_{1}\mu_{B}B+Z_{4}\frac{e^{2}B^{2}R^{2}}{2m}\right)\sigma^{z}.\label{eq_H_L1}
\end{eqnarray}
where $\sigma^{x}$ and $\sigma^{z}$ are Pauli matrices defined in this Hilbert subspace. In the Hilbert subspace spanned by $|e-\rangle$ and $|g+\rangle$, we have the other effective two-band Hamiltonian 
\begin{eqnarray}
H^{\rm ef}&=&\frac{\hbar^{2}k^{2}_{z}}{2m^{*}_{h}}+\left(C\frac{\hbar^{2}}{mR}-Z_{3}\mu_{B}BR\right)k_{z}\sigma^{x}\nonumber\\
&&+\left(\frac{\Delta}{2}\frac{\hbar^{2}}{mR^{2}}-Z_{1}\mu_{B}B+Z_{4}\frac{e^{2}B^{2}R^{2}}{2m}\right)\sigma^{z}.\label{eq_H_L2}
\end{eqnarray}
Hence, we have recovered our previous conclusion that each subset of subband dispersions can be modeled by Eq.~(\ref{eq_dispersion}). These two subsets of subband dispersions are now given by Eqs.~(\ref{eq_H_L1}) and (\ref{eq_H_L2}), respectively. Also, the magnetic field dependent `spin'-orbit coupling $\alpha(B)$ and effective $g$-factor $g^{*}_{h}(B)$ now have explicit expressions. Note that the effective hole mass $m^{*}_{h}$ here does not depend on the magnetic field, this result consists with the previous band fitting result where $m^{*}_{h}$ only has a very small magnetic field dependence~\cite{RL2022b}.

\begin{figure}
\includegraphics{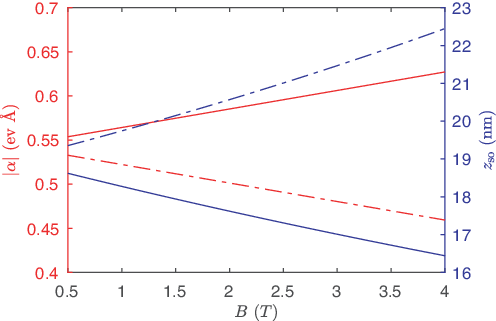}
\caption{\label{fig_soc_l}Both the `spin'-orbit coupling $\alpha$ and the `spin'-orbit length $z_{\rm so}=\hbar^{2}/(m^{*}_{h}\alpha)$ as a function of the longitudinal field. The results of the lowest [Eq.~(\ref{eq_H_L2})] and the second lowest [Eq.~(\ref{eq_H_L1})] subband dispersions are given by the solid and the dashed-dot lines, respectively. Here we have used the band fitting parameters $m^{*}_{h}=0.074m$, $C=7.12$, and $Z_{3}=-3.62$.}
\end{figure}

The hole `spin'-orbit coupling $\alpha(B)$ has a simple linear dependence on the longitudinal field. It increases with the field in one subband dispersion, while it decreases with field in the other subband dispersion (see Fig.~\ref{fig_soc_l}). We also show in Fig.~\ref{fig_soc_l} the magnetic field dependence of the `spin'-orbit length $z_{\rm so}=\hbar^{2}/(m^{*}_{h}\alpha)$, which is another useful measure of `spin'-orbit coupling in the nanowire.  Note that the `spin'-orbit coupling $\alpha$ has an order of several fractions of eV \AA, and the `spin'-orbit length $z_{\rm so}$ has an order of $20$ nm. 

\begin{figure}
\includegraphics{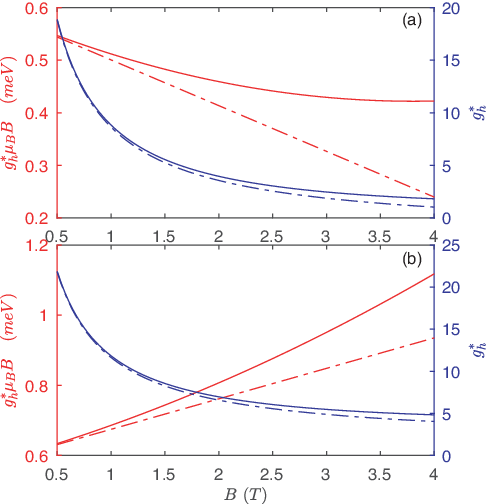}
\caption{\label{fig_gL}Both the Zeeman splitting and the effective $g$-factor $g^{*}_{h}$ of the hole `spin' as a function of the longitudinal field. (a) The result of the lowest [Eq.~(\ref{eq_H_L2})] subband dispersion. (b) The result of the second lowest [Eq.~(\ref{eq_H_L1})] subband dispersion. For comparison, we also give the corresponding results (the dashed dot lines) when the $Z_{4}$ term is neglected.}
\end{figure}

We show in Fig.~\ref{fig_gL} the effects of the $Z_{4}$ term on both the Zeeman splitting and the effective $g$-factor $g^{*}_{h}$ of the hole `spin'. Figures~\ref{fig_gL}(a) and (b) show the results of the lowest [Eq.~(\ref{eq_H_L2})] and second lowest [Eq.~(\ref{eq_H_L1})] subband dispersions, respectively. As expected, the $Z_{4}$ term is negligible in the weak field region, because it is proportional to $B^{2}$. The $Z_{4}$ term becomes important in the strong field region, we can see the obvious difference between the solid line (with $Z_{4}$ term) and dashed-dot line (without $Z_{4}$ term) in the Zeeman splitting of the hole `spin' when $B$ is large. The effects of $Z_{4}$ term on the $g$-factor of the hole `spin' is relatively less obvious. Note that in the weak field region, the $g$-factor is very large, e.g., $g^{*}_{h}\approx20$ at $B=0.5$ T, such that the hole `spin' is a very suitable quantum information carrier for quantum computing~\cite{trif2008spin,nadj2010spin,RL2013,RL2018c,RL2020}.

\subsection{The strong transverse field case}
 
When the magnetic field is applied transversely, i.e., perpendicular to the nanowire, the effective Hamiltonian $H_{t}=H-H_{0}$ of the magnetic terms can be written as
\begin{equation}
H^{\rm ef}_{t}=\mu_{B}B(X_{1}s^{x}+X_{2}\tau^{z}s^{x}+X_{3}k_{z}R\tau^{x})+X_{4}\frac{e^{2}B^{2}R^{2}}{2m}\tau^{z},
\end{equation}
where the $X_{i}$ ($i=1,2,3,4$) terms are introduced to describe both the Zeeman and the orbital effects of the transverse magnetic field. The $X_{1,2,3}$ terms have also been given in Ref.~\cite{PhysRevB.84.195314}, and the $X_{4}$ term is induced by terms proportional to $B^{2}$. The realistic values of $X_{1,2,3,4}$ are given in Tab.~\ref{tab1}. Now, it is more obvious to see the block diagonalized feather of the total Hamiltonian $H^{\rm ef}=H^{\rm ef}_{0}+H^{\rm ef}_{t}$ than that in the longitudinal field case, because the operator $s^{x}$ is a conserved quantity here. For $s^{x}=1$, the total effective Hamiltonian $H^{\rm ef}$ reads
\begin{eqnarray}
H^{\rm ef}&=&\frac{\hbar^{2}k^{2}_{z}}{2m^{*}_{h}}+\left(C\frac{\hbar^{2}}{mR}+X_{3}\mu_{B}BR\right)k_{z}\tau^{x}\nonumber\\
&&+\left(\frac{\Delta}{2}\frac{\hbar^{2}}{mR^{2}}+X_{2}\mu_{B}B+X_{4}\frac{e^{2}B^{2}R^{2}}{2m}\right)\tau^{z}.\label{eq_H_T1}
\end{eqnarray}
For $s^{x}=-1$, the total effective Hamiltonian $H^{\rm ef}$ reads
\begin{eqnarray}
H^{\rm ef}&=&\frac{\hbar^{2}k^{2}_{z}}{2m^{*}_{h}}-\left(C\frac{\hbar^{2}}{mR}-X_{3}\mu_{B}BR\right)k_{z}\tau^{x}\nonumber\\
&&+\left(\frac{\Delta}{2}\frac{\hbar^{2}}{mR^{2}}-X_{2}\mu_{B}B+X_{4}\frac{e^{2}B^{2}R^{2}}{2m}\right)\tau^{z}.\label{eq_H_T2}
\end{eqnarray}
Therefore, when the magnetic field is transverse, we have still recovered Eq.~(\ref{eq_dispersion}) given previously. The two subsets of subband dispersions are now given by Eqs.~(\ref{eq_H_T1}) and (\ref{eq_H_T2}), respectively. The magnetic field dependent `spin'-orbit coupling $\alpha(B)$ and effective $g$-factor $g^{*}_{h}(B)$ also have explicit expressions, just as the longitudinal field case.

\begin{figure}
\includegraphics{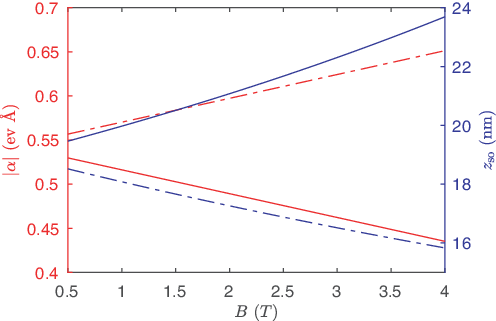}
\caption{\label{fig_soc_t}Both the `spin'-orbit coupling $\alpha$ and the `spin'-orbit length $z_{\rm so}=\hbar^{2}/(m^{*}_{h}\alpha)$ as a function of the transverse field. The results of the lowest [Eq.~(\ref{eq_H_T2})] and the second lowest [Eq.~(\ref{eq_H_T1})] subband dispersions are given by the solid and the dashed-dot lines, respectively. Here we have used the band fitting parameters $m^{*}_{h}=0.074m$, $C=7.12$, and $X_{3}=4.66$.}
\end{figure}

The transverse field dependences of both the `spin'-orbit coupling $\alpha(B)$ and the `spin'-orbit length $z_{\rm so}=\hbar^{2}/(m^{*}_{h}\alpha)$ are shown in Fig.~\ref{fig_soc_t}. The `spin'-orbit coupling is still linearly dependent on the transverse field. Because the parameter $|X_{3}|\approx4.66$ in a transverse field is larger than the parameter $|Z_{3}|=3.62$ in a longitudinal field, it is relatively easier to tune the `spin'-orbit coupling via a transverse field. Note that the order of the `spin'-orbit coupling $\alpha$ is about several fractions of eV \AA, and the order of the `spin'-orbit length $z_{\rm so}$ is about $20$ nm.

\begin{figure}
\includegraphics{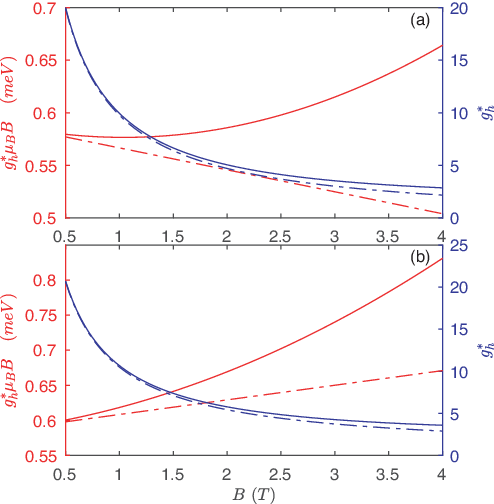}
\caption{\label{fig_gT}Both the Zeeman splitting and the effective $g$-factor $g^{*}_{h}$ of the hole `spin' as a function of the transverse field. (a) The result of the second lowest [Eq.~(\ref{eq_H_T1})] subband dispersion. (b) The result of the lowest [Eq.~(\ref{eq_H_T2})] subband dispersion. For comparison, we also give the corresponding results (the dashed dot lines) when the $X_{4}$ term is neglected.}
\end{figure}

We show in Fig.~\ref{fig_gT} the effects of the $X_{4}$ term on both the Zeeman splitting and the effective $g$-factor $g^{*}_{h}$ of the hole `spin'.  Figures~\ref{fig_gT}(a) and (b) show the results of the second lowest [Eq.~(\ref{eq_H_T1})] and the lowest [Eq.~(\ref{eq_H_T2})] subband dispersions, respectively. Like the case of the longitudinal field, the effects of the $X_{4}$ term on the Zeeman splitting is still more obvious than that on the effective $g$-factor. Especially, because $X_{2}\approx-0.18$ is relatively small here, the Zeeman splitting $g_{e}\mu_{B}B$ in Eq.~(\ref{eq_H_T1}) becomes as an increasing function of $B$ when $B>1.04$ T [see Fig.~\ref{fig_gT}(a)]. Note that, although each subset of subband dispersions is accurately described by an effective two-band Hamiltonian, these two subsets are not well separated from each other in the energy scale [see Fig.~\ref{fig_subbands}(b)]. This result may limit some potential applications of the hole gas in a strong transverse field.

\section{Hole subband dispersions in a general field}

\begin{figure}
\includegraphics{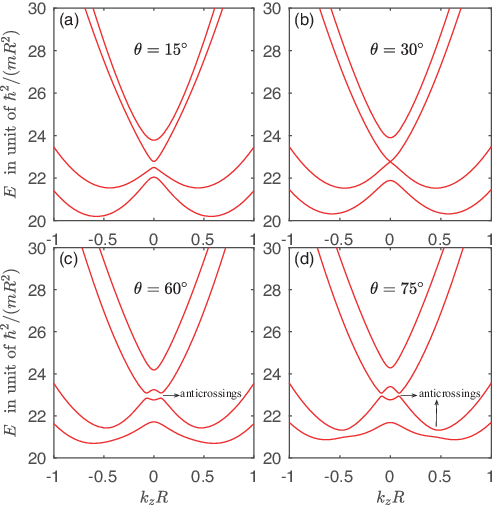}
\caption{\label{fig_angle}Subband dispersions of the hole gas in a general magnetic field. The field strength is fixed at 4 T. The results for magnetic field azimuth $\theta=15^{\circ}$ (a), $30^{\circ}$ (b), $60^{\circ}$ (c), and $75^{\circ}$ (d). There are subband anticrossings (see the sites marked by the arrows), which imply we cannot label the dispersion lines with definite spin index. }
\end{figure}

In the previous sections, we have shown both numerically and analytically that the hole subband dispersions in a longitudinal field or transverse field can be classified using definite hole spin index. One subset of subband dispersions belongs to spin up and the other subset belongs to spin down. Now, we move to study whether this classification is still feasible when the magnetic field is applied in a general direction. We use $\theta$ to denote the azimuth of the magnetic field, i.e., the angle between the magnetic field and the nanowire axis. The general magnetic field can be written as ${\bf B}=(B\sin\theta,0,B\cos\theta)$, and the vector potential can be conveniently chosen as ${\bf A}=(-\frac{1}{2}yB\cos\theta,\frac{1}{2}xB\cos\theta,yB\sin\theta)$. Now, the operator $p_{z}=\hbar\,k_{z}$ in Hamiltonian (\ref{eq_model}) is still a conserved quantity, and the subband quantization still gives rise to $E_{n}(k_{z})$.

By treating all the magnetic terms as a perturbation~\cite{RL2022a}, we obtain the subband dispersions of the hole gas for various magnetic field directions (see Fig.~\ref{fig_angle} with a fixed field strength $B=4$ T). Instead of the subband crossings in a longitudinal field or transverse field (see Fig.~\ref{fig_subbands}), now there are subband anticrossings [see Figs.~\ref{fig_angle}(c) and (d)]. These subband anticrossings indicate that we cannot exactly classify the subband dispersions using the hole spin index. However, when the gaps given at the anticrossings are small enough in comparison with the energy gap $\Delta$ given at $k_{z}R=0$, such as those shown in Figs.~\ref{fig_angle}(c) and (d), we can approximate these anticrossings just as crossings, such that each of the two double-well structures shown in Figs.~\ref{fig_angle}(c) and (d) can still be modeled by Eq.~(\ref{eq_dispersion}).

When the angle of the magnetic field is slowly varied from $0^{\circ}$ to $90^{\circ}$, the subband dispersion I (II) in Fig.~\ref{fig_subbands}(a) will smoothly convert to the dispersion I (II) in Fig.~\ref{fig_subbands}(b). This picture of transition can help to understand the variation tendency of the subband dispersions shown in Fig.~\ref{fig_angle} when $\theta$ varies. Let us focus on the anticrossing near the site $k_{z}R\approx0.5$ shown in Fig.~\ref{fig_angle}(d). Note that the dispersion II is completely above the dispersion I at $\theta=0^{\circ}$ [see Fig.~\ref{fig_subbands}(a)], while these two dispersions become crossing at $90^{\circ}$ [see Fig.~\ref{fig_subbands}(b)], such that there must be an angle where the anticrossing begins and finally becomes as a crossing at $90^{\circ}$. By analyzing our data, this anticrossing roughly begins at $\theta\approx65^{\circ}$. For $\theta>75^{\circ}$, the gap at this anticrossing can be considered as small, and for $\theta<65^{\circ}$, there is no anticrossing.

In an accurate manner, we should use the following effective Hamiltonian
\begin{equation}
H^{\rm ef}= H^{\rm ef}_{0}+\left.H^{\rm ef}_{l}\right|_{B=B\cos\theta}+\left.H^{\rm ef}_{t}\right|_{B=B\sin\theta},
\end{equation} 
to describe the hole subband dispersions in a general magnetic field. This Hamiltonian is certainly a four-band Hamiltonian, we are no longer to exactly rewrite this Hamiltonian as a $2\times2$ block diagonalized Hamiltonian. The two-band description given by Eq.~(\ref{eq_dispersion}) is only an approximation, which is expected to be a good approximation when the band gaps given at the anticrossings are small enough.

\section{Discussion and summary}
We emphasize that the hole `spin' studied in this paper is essentially different from the real hole spin. The real hole spin splitting in a magnetic field is the splitting between the lines I and II shown in Fig.~\ref{fig_subbands}. While the hole `spin' is introduced to describe the two labeled I (II) lines. Although the `spin'-orbit coupling studied here has the same form as that of the Rashba spin-orbit coupling induced by an external electric field~\cite{PhysRevB.84.195314,PhysRevLett.119.126401}, their underlying physics are totally different. The strong `spin'-orbit coupling is completely due to the peculiar double-well anticrossing structure in the hole subband dispersions of the Ge nanowire~\cite{PhysRevB.84.195314,RL2021}. Also, recent electronic structure calculations in InSb and GaSb nanowires with a square cross section have shown similar double-well anticrossing structure~\cite{doi:10.1063/1.4929412}.

In summary, by using the Luttinger-Kohn Hamiltonian in the spherical approximation, we show the subband dispersions of the 1D hole gas in a strong longitudinal or transverse fields can be accurately divided into two subsets. Each subset is described by an effective two-band Hamiltonian where there is a strong hole `spin'-orbit coupling. If the `spin'-orbit coupling in one subset increases with the magnetic field, then the `spin'-orbit coupling in the other subset must decease with magnetic field. We also found there is a $Z_{4}/X_{4}$ term in a strong longitudinal/transverse field that contributes significantly to the Zeeman splitting of the hole `spin'. When the magnetic field is applied in an arbitrary direction, the hole subband dispersions are no-longer exactly divided into two subsets, the two-band Hamiltonian description now is only a good approximation.

\section*{Acknowledgements}
This work was supported by the National Natural Science Foundation of China Grant No.~11404020, the Project from the Department of Education of Hebei Province Grant No. QN2019057, and the Starting up Foundation from Yanshan University Grant No. BL18043.

\bibliography{Ref_Hole_spin}

\begin{thebibliography}{59}%
\makeatletter
\providecommand \@ifxundefined [1]{%
 \@ifx{#1\undefined}
}%
\providecommand \@ifnum [1]{%
 \ifnum #1\expandafter \@firstoftwo
 \else \expandafter \@secondoftwo
 \fi
}%
\providecommand \@ifx [1]{%
 \ifx #1\expandafter \@firstoftwo
 \else \expandafter \@secondoftwo
 \fi
}%
\providecommand \natexlab [1]{#1}%
\providecommand \enquote  [1]{``#1''}%
\providecommand \bibnamefont  [1]{#1}%
\providecommand \bibfnamefont [1]{#1}%
\providecommand \citenamefont [1]{#1}%
\providecommand \href@noop [0]{\@secondoftwo}%
\providecommand \href [0]{\begingroup \@sanitize@url \@href}%
\providecommand \@href[1]{\@@startlink{#1}\@@href}%
\providecommand \@@href[1]{\endgroup#1\@@endlink}%
\providecommand \@sanitize@url [0]{\catcode `\\12\catcode `\$12\catcode
  `\&12\catcode `\#12\catcode `\^12\catcode `\_12\catcode `\%12\relax}%
\providecommand \@@startlink[1]{}%
\providecommand \@@endlink[0]{}%
\providecommand \url  [0]{\begingroup\@sanitize@url \@url }%
\providecommand \@url [1]{\endgroup\@href {#1}{\urlprefix }}%
\providecommand \urlprefix  [0]{URL }%
\providecommand \Eprint [0]{\href }%
\providecommand \doibase [0]{https://doi.org/}%
\providecommand \selectlanguage [0]{\@gobble}%
\providecommand \bibinfo  [0]{\@secondoftwo}%
\providecommand \bibfield  [0]{\@secondoftwo}%
\providecommand \translation [1]{[#1]}%
\providecommand \BibitemOpen [0]{}%
\providecommand \bibitemStop [0]{}%
\providecommand \bibitemNoStop [0]{.\EOS\space}%
\providecommand \EOS [0]{\spacefactor3000\relax}%
\providecommand \BibitemShut  [1]{\csname bibitem#1\endcsname}%
\let\auto@bib@innerbib\@empty
\bibitem [{\citenamefont {Luttinger}\ and\ \citenamefont
  {Kohn}(1955)}]{PhysRev.97.869}%
  \BibitemOpen
  \bibfield  {author} {\bibinfo {author} {\bibfnamefont {J.~M.}\ \bibnamefont
  {Luttinger}}\ and\ \bibinfo {author} {\bibfnamefont {W.}~\bibnamefont
  {Kohn}},\ }\bibfield  {title} {\bibinfo {title} {Motion of electrons and
  holes in perturbed periodic fields},\ }\href
  {https://doi.org/10.1103/PhysRev.97.869} {\bibfield  {journal} {\bibinfo
  {journal} {Phys. Rev.}\ }\textbf {\bibinfo {volume} {97}},\ \bibinfo {pages}
  {869} (\bibinfo {year} {1955})}\BibitemShut {NoStop}%
\bibitem [{\citenamefont {Luttinger}(1956)}]{PhysRev.102.1030}%
  \BibitemOpen
  \bibfield  {author} {\bibinfo {author} {\bibfnamefont {J.~M.}\ \bibnamefont
  {Luttinger}},\ }\bibfield  {title} {\bibinfo {title} {Quantum theory of
  cyclotron resonance in semiconductors: General theory},\ }\href
  {https://doi.org/10.1103/PhysRev.102.1030} {\bibfield  {journal} {\bibinfo
  {journal} {Phys. Rev.}\ }\textbf {\bibinfo {volume} {102}},\ \bibinfo {pages}
  {1030} (\bibinfo {year} {1956})}\BibitemShut {NoStop}%
\bibitem [{\citenamefont {Broido}\ and\ \citenamefont
  {Sham}(1985)}]{PhysRevB.31.888}%
  \BibitemOpen
  \bibfield  {author} {\bibinfo {author} {\bibfnamefont {D.~A.}\ \bibnamefont
  {Broido}}\ and\ \bibinfo {author} {\bibfnamefont {L.~J.}\ \bibnamefont
  {Sham}},\ }\bibfield  {title} {\bibinfo {title} {Effective masses of holes at
  {G}a{A}s-{A}l{G}a{A}s heterojunctions},\ }\href
  {https://doi.org/10.1103/PhysRevB.31.888} {\bibfield  {journal} {\bibinfo
  {journal} {Phys. Rev. B}\ }\textbf {\bibinfo {volume} {31}},\ \bibinfo
  {pages} {888} (\bibinfo {year} {1985})}\BibitemShut {NoStop}%
\bibitem [{\citenamefont {Chuang}(1991)}]{PhysRevB.43.9649}%
  \BibitemOpen
  \bibfield  {author} {\bibinfo {author} {\bibfnamefont {S.~L.}\ \bibnamefont
  {Chuang}},\ }\bibfield  {title} {\bibinfo {title} {Efficient band-structure
  calculations of strained quantum wells},\ }\href
  {https://doi.org/10.1103/PhysRevB.43.9649} {\bibfield  {journal} {\bibinfo
  {journal} {Phys. Rev. B}\ }\textbf {\bibinfo {volume} {43}},\ \bibinfo
  {pages} {9649} (\bibinfo {year} {1991})}\BibitemShut {NoStop}%
\bibitem [{\citenamefont {Marcellina}\ \emph {et~al.}(2017)\citenamefont
  {Marcellina}, \citenamefont {Hamilton}, \citenamefont {Winkler},\ and\
  \citenamefont {Culcer}}]{PhysRevB.95.075305}%
  \BibitemOpen
  \bibfield  {author} {\bibinfo {author} {\bibfnamefont {E.}~\bibnamefont
  {Marcellina}}, \bibinfo {author} {\bibfnamefont {A.~R.}\ \bibnamefont
  {Hamilton}}, \bibinfo {author} {\bibfnamefont {R.}~\bibnamefont {Winkler}},\
  and\ \bibinfo {author} {\bibfnamefont {D.}~\bibnamefont {Culcer}},\
  }\bibfield  {title} {\bibinfo {title} {Spin-orbit interactions in
  inversion-asymmetric two-dimensional hole systems: A variational analysis},\
  }\href {https://doi.org/10.1103/PhysRevB.95.075305} {\bibfield  {journal}
  {\bibinfo  {journal} {Phys. Rev. B}\ }\textbf {\bibinfo {volume} {95}},\
  \bibinfo {pages} {075305} (\bibinfo {year} {2017})}\BibitemShut {NoStop}%
\bibitem [{\citenamefont {Xia}(1989)}]{PhysRevB.40.8500}%
  \BibitemOpen
  \bibfield  {author} {\bibinfo {author} {\bibfnamefont {J.-B.}\ \bibnamefont
  {Xia}},\ }\bibfield  {title} {\bibinfo {title} {Electronic structures of
  zero-dimensional quantum wells},\ }\href
  {https://doi.org/10.1103/PhysRevB.40.8500} {\bibfield  {journal} {\bibinfo
  {journal} {Phys. Rev. B}\ }\textbf {\bibinfo {volume} {40}},\ \bibinfo
  {pages} {8500} (\bibinfo {year} {1989})}\BibitemShut {NoStop}%
\bibitem [{\citenamefont {Baldereschi}\ and\ \citenamefont
  {Lipari}(1973)}]{PhysRevB.8.2697}%
  \BibitemOpen
  \bibfield  {author} {\bibinfo {author} {\bibfnamefont {A.}~\bibnamefont
  {Baldereschi}}\ and\ \bibinfo {author} {\bibfnamefont {N.~O.}\ \bibnamefont
  {Lipari}},\ }\bibfield  {title} {\bibinfo {title} {Spherical model of shallow
  acceptor states in semiconductors},\ }\href
  {https://doi.org/10.1103/PhysRevB.8.2697} {\bibfield  {journal} {\bibinfo
  {journal} {Phys. Rev. B}\ }\textbf {\bibinfo {volume} {8}},\ \bibinfo {pages}
  {2697} (\bibinfo {year} {1973})}\BibitemShut {NoStop}%
\bibitem [{\citenamefont {Semina}\ and\ \citenamefont
  {Suris}(2015)}]{Semina:2015aa}%
  \BibitemOpen
  \bibfield  {author} {\bibinfo {author} {\bibfnamefont {M.~A.}\ \bibnamefont
  {Semina}}\ and\ \bibinfo {author} {\bibfnamefont {R.~A.}\ \bibnamefont
  {Suris}},\ }\bibfield  {title} {\bibinfo {title} {Holes localized in
  nanostructures in an external magnetic field: g-factor and mixing of
  states},\ }\href {https://doi.org/10.1134/S1063782615060214} {\bibfield
  {journal} {\bibinfo  {journal} {Semiconductors}\ }\textbf {\bibinfo {volume}
  {49}},\ \bibinfo {pages} {797} (\bibinfo {year} {2015})}\BibitemShut
  {NoStop}%
\bibitem [{\citenamefont {Wu}\ \emph {et~al.}(2010)\citenamefont {Wu},
  \citenamefont {Jiang},\ and\ \citenamefont {Weng}}]{WU201061}%
  \BibitemOpen
  \bibfield  {author} {\bibinfo {author} {\bibfnamefont {M.}~\bibnamefont
  {Wu}}, \bibinfo {author} {\bibfnamefont {J.}~\bibnamefont {Jiang}},\ and\
  \bibinfo {author} {\bibfnamefont {M.}~\bibnamefont {Weng}},\ }\bibfield
  {title} {\bibinfo {title} {Spin dynamics in semiconductors},\ }\href
  {https://doi.org/https://doi.org/10.1016/j.physrep.2010.04.002} {\bibfield
  {journal} {\bibinfo  {journal} {Physics Reports}\ }\textbf {\bibinfo {volume}
  {493}},\ \bibinfo {pages} {61} (\bibinfo {year} {2010})}\BibitemShut
  {NoStop}%
\bibitem [{\citenamefont {Sercel}\ and\ \citenamefont
  {Vahala}(1990)}]{PhysRevB.42.3690}%
  \BibitemOpen
  \bibfield  {author} {\bibinfo {author} {\bibfnamefont {P.~C.}\ \bibnamefont
  {Sercel}}\ and\ \bibinfo {author} {\bibfnamefont {K.~J.}\ \bibnamefont
  {Vahala}},\ }\bibfield  {title} {\bibinfo {title} {Analytical formalism for
  determining quantum-wire and quantum-dot band structure in the multiband
  envelope-function approximation},\ }\href
  {https://doi.org/10.1103/PhysRevB.42.3690} {\bibfield  {journal} {\bibinfo
  {journal} {Phys. Rev. B}\ }\textbf {\bibinfo {volume} {42}},\ \bibinfo
  {pages} {3690} (\bibinfo {year} {1990})}\BibitemShut {NoStop}%
\bibitem [{\citenamefont {Sweeny}\ \emph {et~al.}(1988)\citenamefont {Sweeny},
  \citenamefont {Xu},\ and\ \citenamefont {Shur}}]{sweeny1988hole}%
  \BibitemOpen
  \bibfield  {author} {\bibinfo {author} {\bibfnamefont {M.}~\bibnamefont
  {Sweeny}}, \bibinfo {author} {\bibfnamefont {J.}~\bibnamefont {Xu}},\ and\
  \bibinfo {author} {\bibfnamefont {M.}~\bibnamefont {Shur}},\ }\bibfield
  {title} {\bibinfo {title} {Hole subbands in one-dimensional quantum well
  wires},\ }\href {https://doi.org/10.1016/0749-6036(88)90249-2} {\bibfield
  {journal} {\bibinfo  {journal} {Superlattices and Microstructures}\ }\textbf
  {\bibinfo {volume} {4}},\ \bibinfo {pages} {623} (\bibinfo {year}
  {1988})}\BibitemShut {NoStop}%
\bibitem [{\citenamefont {Andreani}\ \emph {et~al.}(1987)\citenamefont
  {Andreani}, \citenamefont {Pasquarello},\ and\ \citenamefont
  {Bassani}}]{PhysRevB.36.5887}%
  \BibitemOpen
  \bibfield  {author} {\bibinfo {author} {\bibfnamefont {L.~C.}\ \bibnamefont
  {Andreani}}, \bibinfo {author} {\bibfnamefont {A.}~\bibnamefont
  {Pasquarello}},\ and\ \bibinfo {author} {\bibfnamefont {F.}~\bibnamefont
  {Bassani}},\ }\bibfield  {title} {\bibinfo {title} {Hole subbands in strained
  {G}a{A}s-{G}a$_{\rm 1-x}${A}l$_{\rm x}${A}s quantum wells: Exact solution of
  the effective-mass equation},\ }\href
  {https://doi.org/10.1103/PhysRevB.36.5887} {\bibfield  {journal} {\bibinfo
  {journal} {Phys. Rev. B}\ }\textbf {\bibinfo {volume} {36}},\ \bibinfo
  {pages} {5887} (\bibinfo {year} {1987})}\BibitemShut {NoStop}%
\bibitem [{\citenamefont {Ando}(1985)}]{ando1985hole}%
  \BibitemOpen
  \bibfield  {author} {\bibinfo {author} {\bibfnamefont {T.}~\bibnamefont
  {Ando}},\ }\bibfield  {title} {\bibinfo {title} {Hole subband at
  {G}a{A}s/{A}l{G}a{A}s heterojunctions and quantum wells},\ }\href@noop {}
  {\bibfield  {journal} {\bibinfo  {journal} {Journal of the physical society
  of Japan}\ }\textbf {\bibinfo {volume} {54}},\ \bibinfo {pages} {1528}
  (\bibinfo {year} {1985})}\BibitemShut {NoStop}%
\bibitem [{\citenamefont {Rashba}\ and\ \citenamefont
  {Sherman}(1988)}]{RASHBA1988175}%
  \BibitemOpen
  \bibfield  {author} {\bibinfo {author} {\bibfnamefont {E.}~\bibnamefont
  {Rashba}}\ and\ \bibinfo {author} {\bibfnamefont {E.}~\bibnamefont
  {Sherman}},\ }\bibfield  {title} {\bibinfo {title} {Spin-orbital band
  splitting in symmetric quantum wells},\ }\href
  {https://doi.org/https://doi.org/10.1016/0375-9601(88)90140-5} {\bibfield
  {journal} {\bibinfo  {journal} {Physics Letters A}\ }\textbf {\bibinfo
  {volume} {129}},\ \bibinfo {pages} {175} (\bibinfo {year}
  {1988})}\BibitemShut {NoStop}%
\bibitem [{\citenamefont {Fishman}(1995)}]{PhysRevB.52.11132}%
  \BibitemOpen
  \bibfield  {author} {\bibinfo {author} {\bibfnamefont {G.}~\bibnamefont
  {Fishman}},\ }\bibfield  {title} {\bibinfo {title} {Hole subbands in strained
  quantum-well semiconductors in [hhk] directions},\ }\href
  {https://doi.org/10.1103/PhysRevB.52.11132} {\bibfield  {journal} {\bibinfo
  {journal} {Phys. Rev. B}\ }\textbf {\bibinfo {volume} {52}},\ \bibinfo
  {pages} {11132} (\bibinfo {year} {1995})}\BibitemShut {NoStop}%
\bibitem [{\citenamefont {Hendrickx}\ \emph
  {et~al.}(2020{\natexlab{a}})\citenamefont {Hendrickx}, \citenamefont
  {Lawrie}, \citenamefont {Petit}, \citenamefont {Sammak}, \citenamefont
  {Scappucci},\ and\ \citenamefont {Veldhorst}}]{Hendrickx:2020ab}%
  \BibitemOpen
  \bibfield  {author} {\bibinfo {author} {\bibfnamefont {N.~W.}\ \bibnamefont
  {Hendrickx}}, \bibinfo {author} {\bibfnamefont {W.~I.~L.}\ \bibnamefont
  {Lawrie}}, \bibinfo {author} {\bibfnamefont {L.}~\bibnamefont {Petit}},
  \bibinfo {author} {\bibfnamefont {A.}~\bibnamefont {Sammak}}, \bibinfo
  {author} {\bibfnamefont {G.}~\bibnamefont {Scappucci}},\ and\ \bibinfo
  {author} {\bibfnamefont {M.}~\bibnamefont {Veldhorst}},\ }\bibfield  {title}
  {\bibinfo {title} {A single-hole spin qubit},\ }\href
  {https://doi.org/10.1038/s41467-020-17211-7} {\bibfield  {journal} {\bibinfo
  {journal} {Nature Communications}\ }\textbf {\bibinfo {volume} {11}},\
  \bibinfo {pages} {3478} (\bibinfo {year} {2020}{\natexlab{a}})}\BibitemShut
  {NoStop}%
\bibitem [{\citenamefont {Hendrickx}\ \emph
  {et~al.}(2020{\natexlab{b}})\citenamefont {Hendrickx}, \citenamefont
  {Franke}, \citenamefont {Sammak}, \citenamefont {Scappucci},\ and\
  \citenamefont {Veldhorst}}]{Hendrickx:2020aa}%
  \BibitemOpen
  \bibfield  {author} {\bibinfo {author} {\bibfnamefont {N.~W.}\ \bibnamefont
  {Hendrickx}}, \bibinfo {author} {\bibfnamefont {D.~P.}\ \bibnamefont
  {Franke}}, \bibinfo {author} {\bibfnamefont {A.}~\bibnamefont {Sammak}},
  \bibinfo {author} {\bibfnamefont {G.}~\bibnamefont {Scappucci}},\ and\
  \bibinfo {author} {\bibfnamefont {M.}~\bibnamefont {Veldhorst}},\ }\bibfield
  {title} {\bibinfo {title} {Fast two-qubit logic with holes in germanium},\
  }\href {https://doi.org/10.1038/s41586-019-1919-3} {\bibfield  {journal}
  {\bibinfo  {journal} {Nature}\ }\textbf {\bibinfo {volume} {577}},\ \bibinfo
  {pages} {487} (\bibinfo {year} {2020}{\natexlab{b}})}\BibitemShut {NoStop}%
\bibitem [{\citenamefont {Scappucci}\ \emph {et~al.}(2021)\citenamefont
  {Scappucci}, \citenamefont {Kloeffel}, \citenamefont {Zwanenburg},
  \citenamefont {Loss}, \citenamefont {Myronov}, \citenamefont {Zhang},
  \citenamefont {De~Franceschi}, \citenamefont {Katsaros},\ and\ \citenamefont
  {Veldhorst}}]{Scappucci:2021vk}%
  \BibitemOpen
  \bibfield  {author} {\bibinfo {author} {\bibfnamefont {G.}~\bibnamefont
  {Scappucci}}, \bibinfo {author} {\bibfnamefont {C.}~\bibnamefont {Kloeffel}},
  \bibinfo {author} {\bibfnamefont {F.~A.}\ \bibnamefont {Zwanenburg}},
  \bibinfo {author} {\bibfnamefont {D.}~\bibnamefont {Loss}}, \bibinfo {author}
  {\bibfnamefont {M.}~\bibnamefont {Myronov}}, \bibinfo {author} {\bibfnamefont
  {J.-J.}\ \bibnamefont {Zhang}}, \bibinfo {author} {\bibfnamefont
  {S.}~\bibnamefont {De~Franceschi}}, \bibinfo {author} {\bibfnamefont
  {G.}~\bibnamefont {Katsaros}},\ and\ \bibinfo {author} {\bibfnamefont
  {M.}~\bibnamefont {Veldhorst}},\ }\bibfield  {title} {\bibinfo {title} {The
  germanium quantum information route},\ }\href
  {https://doi.org/10.1038/s41578-020-00262-z} {\bibfield  {journal} {\bibinfo
  {journal} {Nature Reviews Materials}\ }\textbf {\bibinfo {volume} {6}},\
  \bibinfo {pages} {926} (\bibinfo {year} {2021})}\BibitemShut {NoStop}%
\bibitem [{\citenamefont {Winkler}(2003)}]{winkler2003spin}%
  \BibitemOpen
  \bibfield  {author} {\bibinfo {author} {\bibfnamefont {R.}~\bibnamefont
  {Winkler}},\ }\href@noop {} {\emph {\bibinfo {title} {Spin-Orbit Effects in
  Two-Dimensional Electron and Hole Systems}}}\ (\bibinfo  {publisher}
  {Springer, Berlin},\ \bibinfo {year} {2003})\BibitemShut {NoStop}%
\bibitem [{\citenamefont {Winkler}(2000)}]{PhysRevB.62.4245}%
  \BibitemOpen
  \bibfield  {author} {\bibinfo {author} {\bibfnamefont {R.}~\bibnamefont
  {Winkler}},\ }\bibfield  {title} {\bibinfo {title} {Rashba spin splitting in
  two-dimensional electron and hole systems},\ }\href
  {https://doi.org/10.1103/PhysRevB.62.4245} {\bibfield  {journal} {\bibinfo
  {journal} {Phys. Rev. B}\ }\textbf {\bibinfo {volume} {62}},\ \bibinfo
  {pages} {4245} (\bibinfo {year} {2000})}\BibitemShut {NoStop}%
\bibitem [{\citenamefont {Xiong}\ \emph {et~al.}(2021)\citenamefont {Xiong},
  \citenamefont {Guan}, \citenamefont {Luo},\ and\ \citenamefont
  {Li}}]{PhysRevB.103.085309}%
  \BibitemOpen
  \bibfield  {author} {\bibinfo {author} {\bibfnamefont {J.-X.}\ \bibnamefont
  {Xiong}}, \bibinfo {author} {\bibfnamefont {S.}~\bibnamefont {Guan}},
  \bibinfo {author} {\bibfnamefont {J.-W.}\ \bibnamefont {Luo}},\ and\ \bibinfo
  {author} {\bibfnamefont {S.-S.}\ \bibnamefont {Li}},\ }\bibfield  {title}
  {\bibinfo {title} {Emergence of strong tunable linear rashba spin-orbit
  coupling in two-dimensional hole gases in semiconductor quantum wells},\
  }\href {https://doi.org/10.1103/PhysRevB.103.085309} {\bibfield  {journal}
  {\bibinfo  {journal} {Phys. Rev. B}\ }\textbf {\bibinfo {volume} {103}},\
  \bibinfo {pages} {085309} (\bibinfo {year} {2021})}\BibitemShut {NoStop}%
\bibitem [{\citenamefont {Budkin}\ and\ \citenamefont
  {Tarasenko}(2022)}]{PhysRevB.105.L161301}%
  \BibitemOpen
  \bibfield  {author} {\bibinfo {author} {\bibfnamefont {G.~V.}\ \bibnamefont
  {Budkin}}\ and\ \bibinfo {author} {\bibfnamefont {S.~A.}\ \bibnamefont
  {Tarasenko}},\ }\bibfield  {title} {\bibinfo {title} {Spin splitting in
  low-symmetry quantum wells beyond {Rashba} and {Dresselhaus} terms},\ }\href
  {https://doi.org/10.1103/PhysRevB.105.L161301} {\bibfield  {journal}
  {\bibinfo  {journal} {Phys. Rev. B}\ }\textbf {\bibinfo {volume} {105}},\
  \bibinfo {pages} {L161301} (\bibinfo {year} {2022})}\BibitemShut {NoStop}%
\bibitem [{\citenamefont {Bulaev}\ and\ \citenamefont
  {Loss}(2007)}]{PhysRevLett.98.097202}%
  \BibitemOpen
  \bibfield  {author} {\bibinfo {author} {\bibfnamefont {D.~V.}\ \bibnamefont
  {Bulaev}}\ and\ \bibinfo {author} {\bibfnamefont {D.}~\bibnamefont {Loss}},\
  }\bibfield  {title} {\bibinfo {title} {Electric dipole spin resonance for
  heavy holes in quantum dots},\ }\href
  {https://doi.org/10.1103/PhysRevLett.98.097202} {\bibfield  {journal}
  {\bibinfo  {journal} {Phys. Rev. Lett.}\ }\textbf {\bibinfo {volume} {98}},\
  \bibinfo {pages} {097202} (\bibinfo {year} {2007})}\BibitemShut {NoStop}%
\bibitem [{\citenamefont {Wang}\ \emph {et~al.}(2021)\citenamefont {Wang},
  \citenamefont {Marcellina}, \citenamefont {Hamilton}, \citenamefont {Cullen},
  \citenamefont {Rogge}, \citenamefont {Salfi},\ and\ \citenamefont
  {Culcer}}]{Wang:2021wc}%
  \BibitemOpen
  \bibfield  {author} {\bibinfo {author} {\bibfnamefont {Z.}~\bibnamefont
  {Wang}}, \bibinfo {author} {\bibfnamefont {E.}~\bibnamefont {Marcellina}},
  \bibinfo {author} {\bibfnamefont {A.~R.}\ \bibnamefont {Hamilton}}, \bibinfo
  {author} {\bibfnamefont {J.~H.}\ \bibnamefont {Cullen}}, \bibinfo {author}
  {\bibfnamefont {S.}~\bibnamefont {Rogge}}, \bibinfo {author} {\bibfnamefont
  {J.}~\bibnamefont {Salfi}},\ and\ \bibinfo {author} {\bibfnamefont
  {D.}~\bibnamefont {Culcer}},\ }\bibfield  {title} {\bibinfo {title} {Optimal
  operation points for ultrafast, highly coherent {Ge} hole spin-orbit
  qubits},\ }\href {https://doi.org/10.1038/s41534-021-00386-2} {\bibfield
  {journal} {\bibinfo  {journal} {npj Quantum Information}\ }\textbf {\bibinfo
  {volume} {7}},\ \bibinfo {pages} {54} (\bibinfo {year} {2021})}\BibitemShut
  {NoStop}%
\bibitem [{\citenamefont {Terrazos}\ \emph {et~al.}(2021)\citenamefont
  {Terrazos}, \citenamefont {Marcellina}, \citenamefont {Wang}, \citenamefont
  {Coppersmith}, \citenamefont {Friesen}, \citenamefont {Hamilton},
  \citenamefont {Hu}, \citenamefont {Koiller}, \citenamefont {Saraiva},
  \citenamefont {Culcer},\ and\ \citenamefont {Capaz}}]{PhysRevB.103.125201}%
  \BibitemOpen
  \bibfield  {author} {\bibinfo {author} {\bibfnamefont {L.~A.}\ \bibnamefont
  {Terrazos}}, \bibinfo {author} {\bibfnamefont {E.}~\bibnamefont
  {Marcellina}}, \bibinfo {author} {\bibfnamefont {Z.}~\bibnamefont {Wang}},
  \bibinfo {author} {\bibfnamefont {S.~N.}\ \bibnamefont {Coppersmith}},
  \bibinfo {author} {\bibfnamefont {M.}~\bibnamefont {Friesen}}, \bibinfo
  {author} {\bibfnamefont {A.~R.}\ \bibnamefont {Hamilton}}, \bibinfo {author}
  {\bibfnamefont {X.}~\bibnamefont {Hu}}, \bibinfo {author} {\bibfnamefont
  {B.}~\bibnamefont {Koiller}}, \bibinfo {author} {\bibfnamefont {A.~L.}\
  \bibnamefont {Saraiva}}, \bibinfo {author} {\bibfnamefont {D.}~\bibnamefont
  {Culcer}},\ and\ \bibinfo {author} {\bibfnamefont {R.~B.}\ \bibnamefont
  {Capaz}},\ }\bibfield  {title} {\bibinfo {title} {Theory of hole-spin qubits
  in strained germanium quantum dots},\ }\href
  {https://doi.org/10.1103/PhysRevB.103.125201} {\bibfield  {journal} {\bibinfo
   {journal} {Phys. Rev. B}\ }\textbf {\bibinfo {volume} {103}},\ \bibinfo
  {pages} {125201} (\bibinfo {year} {2021})}\BibitemShut {NoStop}%
\bibitem [{\citenamefont {Bulaev}\ and\ \citenamefont
  {Loss}(2005)}]{PhysRevLett.95.076805}%
  \BibitemOpen
  \bibfield  {author} {\bibinfo {author} {\bibfnamefont {D.~V.}\ \bibnamefont
  {Bulaev}}\ and\ \bibinfo {author} {\bibfnamefont {D.}~\bibnamefont {Loss}},\
  }\bibfield  {title} {\bibinfo {title} {Spin relaxation and decoherence of
  holes in quantum dots},\ }\href
  {https://doi.org/10.1103/PhysRevLett.95.076805} {\bibfield  {journal}
  {\bibinfo  {journal} {Phys. Rev. Lett.}\ }\textbf {\bibinfo {volume} {95}},\
  \bibinfo {pages} {076805} (\bibinfo {year} {2005})}\BibitemShut {NoStop}%
\bibitem [{\citenamefont {Mutter}\ and\ \citenamefont
  {Burkard}(2020)}]{PhysRevB.102.205412}%
  \BibitemOpen
  \bibfield  {author} {\bibinfo {author} {\bibfnamefont {P.~M.}\ \bibnamefont
  {Mutter}}\ and\ \bibinfo {author} {\bibfnamefont {G.}~\bibnamefont
  {Burkard}},\ }\bibfield  {title} {\bibinfo {title} {Cavity control over
  heavy-hole spin qubits in inversion-symmetric crystals},\ }\href
  {https://doi.org/10.1103/PhysRevB.102.205412} {\bibfield  {journal} {\bibinfo
   {journal} {Phys. Rev. B}\ }\textbf {\bibinfo {volume} {102}},\ \bibinfo
  {pages} {205412} (\bibinfo {year} {2020})}\BibitemShut {NoStop}%
\bibitem [{\citenamefont {Roddaro}\ \emph {et~al.}(2008)\citenamefont
  {Roddaro}, \citenamefont {Fuhrer}, \citenamefont {Brusheim}, \citenamefont
  {Fasth}, \citenamefont {Xu}, \citenamefont {Samuelson}, \citenamefont
  {Xiang},\ and\ \citenamefont {Lieber}}]{PhysRevLett.101.186802}%
  \BibitemOpen
  \bibfield  {author} {\bibinfo {author} {\bibfnamefont {S.}~\bibnamefont
  {Roddaro}}, \bibinfo {author} {\bibfnamefont {A.}~\bibnamefont {Fuhrer}},
  \bibinfo {author} {\bibfnamefont {P.}~\bibnamefont {Brusheim}}, \bibinfo
  {author} {\bibfnamefont {C.}~\bibnamefont {Fasth}}, \bibinfo {author}
  {\bibfnamefont {H.~Q.}\ \bibnamefont {Xu}}, \bibinfo {author} {\bibfnamefont
  {L.}~\bibnamefont {Samuelson}}, \bibinfo {author} {\bibfnamefont
  {J.}~\bibnamefont {Xiang}},\ and\ \bibinfo {author} {\bibfnamefont {C.~M.}\
  \bibnamefont {Lieber}},\ }\bibfield  {title} {\bibinfo {title} {Spin states
  of holes in $\mathrm{Ge}/\mathrm{Si}$ nanowire quantum dots},\ }\href
  {https://doi.org/10.1103/PhysRevLett.101.186802} {\bibfield  {journal}
  {\bibinfo  {journal} {Phys. Rev. Lett.}\ }\textbf {\bibinfo {volume} {101}},\
  \bibinfo {pages} {186802} (\bibinfo {year} {2008})}\BibitemShut {NoStop}%
\bibitem [{\citenamefont {Watzinger}\ \emph {et~al.}(2018)\citenamefont
  {Watzinger}, \citenamefont {Kuku{\v c}ka}, \citenamefont {Vuku{\v s}i{\'c}},
  \citenamefont {Gao}, \citenamefont {Wang}, \citenamefont {Sch{\"a}ffler},
  \citenamefont {Zhang},\ and\ \citenamefont {Katsaros}}]{Watzinger:2018aa}%
  \BibitemOpen
  \bibfield  {author} {\bibinfo {author} {\bibfnamefont {H.}~\bibnamefont
  {Watzinger}}, \bibinfo {author} {\bibfnamefont {J.}~\bibnamefont {Kuku{\v
  c}ka}}, \bibinfo {author} {\bibfnamefont {L.}~\bibnamefont {Vuku{\v
  s}i{\'c}}}, \bibinfo {author} {\bibfnamefont {F.}~\bibnamefont {Gao}},
  \bibinfo {author} {\bibfnamefont {T.}~\bibnamefont {Wang}}, \bibinfo {author}
  {\bibfnamefont {F.}~\bibnamefont {Sch{\"a}ffler}}, \bibinfo {author}
  {\bibfnamefont {J.-J.}\ \bibnamefont {Zhang}},\ and\ \bibinfo {author}
  {\bibfnamefont {G.}~\bibnamefont {Katsaros}},\ }\bibfield  {title} {\bibinfo
  {title} {A germanium hole spin qubit},\ }\href
  {https://doi.org/10.1038/s41467-018-06418-4} {\bibfield  {journal} {\bibinfo
  {journal} {Nature Communications}\ }\textbf {\bibinfo {volume} {9}},\
  \bibinfo {pages} {3902} (\bibinfo {year} {2018})}\BibitemShut {NoStop}%
\bibitem [{\citenamefont {Froning}\ \emph
  {et~al.}(2021{\natexlab{a}})\citenamefont {Froning}, \citenamefont
  {Ran\ifmmode \check{c}\else \v{c}\fi{}i\ifmmode~\acute{c}\else \'{c}\fi{}},
  \citenamefont {Het\'enyi}, \citenamefont {Bosco}, \citenamefont {Rehmann},
  \citenamefont {Li}, \citenamefont {Bakkers}, \citenamefont {Zwanenburg},
  \citenamefont {Loss}, \citenamefont {Zumb\"uhl},\ and\ \citenamefont
  {Braakman}}]{PhysRevResearch.3.013081}%
  \BibitemOpen
  \bibfield  {author} {\bibinfo {author} {\bibfnamefont {F.~N.~M.}\
  \bibnamefont {Froning}}, \bibinfo {author} {\bibfnamefont {M.~J.}\
  \bibnamefont {Ran\ifmmode \check{c}\else \v{c}\fi{}i\ifmmode~\acute{c}\else
  \'{c}\fi{}}}, \bibinfo {author} {\bibfnamefont {B.}~\bibnamefont
  {Het\'enyi}}, \bibinfo {author} {\bibfnamefont {S.}~\bibnamefont {Bosco}},
  \bibinfo {author} {\bibfnamefont {M.~K.}\ \bibnamefont {Rehmann}}, \bibinfo
  {author} {\bibfnamefont {A.}~\bibnamefont {Li}}, \bibinfo {author}
  {\bibfnamefont {E.~P. A.~M.}\ \bibnamefont {Bakkers}}, \bibinfo {author}
  {\bibfnamefont {F.~A.}\ \bibnamefont {Zwanenburg}}, \bibinfo {author}
  {\bibfnamefont {D.}~\bibnamefont {Loss}}, \bibinfo {author} {\bibfnamefont
  {D.~M.}\ \bibnamefont {Zumb\"uhl}},\ and\ \bibinfo {author} {\bibfnamefont
  {F.~R.}\ \bibnamefont {Braakman}},\ }\bibfield  {title} {\bibinfo {title}
  {Strong spin-orbit interaction and $g$-factor renormalization of hole spins
  in {G}e/{S}i nanowire quantum dots},\ }\href
  {https://doi.org/10.1103/PhysRevResearch.3.013081} {\bibfield  {journal}
  {\bibinfo  {journal} {Phys. Rev. Research}\ }\textbf {\bibinfo {volume}
  {3}},\ \bibinfo {pages} {013081} (\bibinfo {year}
  {2021}{\natexlab{a}})}\BibitemShut {NoStop}%
\bibitem [{\citenamefont {Higginbotham}\ \emph {et~al.}(2014)\citenamefont
  {Higginbotham}, \citenamefont {Kuemmeth}, \citenamefont {Larsen},
  \citenamefont {Fitzpatrick}, \citenamefont {Yao}, \citenamefont {Yan},
  \citenamefont {Lieber},\ and\ \citenamefont
  {Marcus}}]{PhysRevLett.112.216806}%
  \BibitemOpen
  \bibfield  {author} {\bibinfo {author} {\bibfnamefont {A.~P.}\ \bibnamefont
  {Higginbotham}}, \bibinfo {author} {\bibfnamefont {F.}~\bibnamefont
  {Kuemmeth}}, \bibinfo {author} {\bibfnamefont {T.~W.}\ \bibnamefont
  {Larsen}}, \bibinfo {author} {\bibfnamefont {M.}~\bibnamefont {Fitzpatrick}},
  \bibinfo {author} {\bibfnamefont {J.}~\bibnamefont {Yao}}, \bibinfo {author}
  {\bibfnamefont {H.}~\bibnamefont {Yan}}, \bibinfo {author} {\bibfnamefont
  {C.~M.}\ \bibnamefont {Lieber}},\ and\ \bibinfo {author} {\bibfnamefont
  {C.~M.}\ \bibnamefont {Marcus}},\ }\bibfield  {title} {\bibinfo {title}
  {Antilocalization of coulomb blockade in a {G}e/{S}i nanowire},\ }\href
  {https://doi.org/10.1103/PhysRevLett.112.216806} {\bibfield  {journal}
  {\bibinfo  {journal} {Phys. Rev. Lett.}\ }\textbf {\bibinfo {volume} {112}},\
  \bibinfo {pages} {216806} (\bibinfo {year} {2014})}\BibitemShut {NoStop}%
\bibitem [{\citenamefont {Froning}\ \emph
  {et~al.}(2021{\natexlab{b}})\citenamefont {Froning}, \citenamefont
  {Camenzind}, \citenamefont {van~der Molen}, \citenamefont {Li}, \citenamefont
  {Bakkers}, \citenamefont {Zumb{\"u}hl},\ and\ \citenamefont
  {Braakman}}]{Froning:2021aa}%
  \BibitemOpen
  \bibfield  {author} {\bibinfo {author} {\bibfnamefont {F.~N.~M.}\
  \bibnamefont {Froning}}, \bibinfo {author} {\bibfnamefont {L.~C.}\
  \bibnamefont {Camenzind}}, \bibinfo {author} {\bibfnamefont {O.~A.~H.}\
  \bibnamefont {van~der Molen}}, \bibinfo {author} {\bibfnamefont
  {A.}~\bibnamefont {Li}}, \bibinfo {author} {\bibfnamefont {E.~P. A.~M.}\
  \bibnamefont {Bakkers}}, \bibinfo {author} {\bibfnamefont {D.~M.}\
  \bibnamefont {Zumb{\"u}hl}},\ and\ \bibinfo {author} {\bibfnamefont {F.~R.}\
  \bibnamefont {Braakman}},\ }\bibfield  {title} {\bibinfo {title} {Ultrafast
  hole spin qubit with gate-tunable spin--orbit switch functionality},\ }\href
  {https://doi.org/10.1038/s41565-020-00828-6} {\bibfield  {journal} {\bibinfo
  {journal} {Nature Nanotechnology}\ }\textbf {\bibinfo {volume} {16}},\
  \bibinfo {pages} {308} (\bibinfo {year} {2021}{\natexlab{b}})}\BibitemShut
  {NoStop}%
\bibitem [{\citenamefont {Wang}\ \emph {et~al.}(2022)\citenamefont {Wang},
  \citenamefont {Xu}, \citenamefont {Gao}, \citenamefont {Liu}, \citenamefont
  {Ma}, \citenamefont {Zhang}, \citenamefont {Wang}, \citenamefont {Cao},
  \citenamefont {Wang}, \citenamefont {Zhang}, \citenamefont {Culcer},
  \citenamefont {Hu}, \citenamefont {Jiang}, \citenamefont {Li}, \citenamefont
  {Guo},\ and\ \citenamefont {Guo}}]{Wang:2022tm}%
  \BibitemOpen
  \bibfield  {author} {\bibinfo {author} {\bibfnamefont {K.}~\bibnamefont
  {Wang}}, \bibinfo {author} {\bibfnamefont {G.}~\bibnamefont {Xu}}, \bibinfo
  {author} {\bibfnamefont {F.}~\bibnamefont {Gao}}, \bibinfo {author}
  {\bibfnamefont {H.}~\bibnamefont {Liu}}, \bibinfo {author} {\bibfnamefont
  {R.-L.}\ \bibnamefont {Ma}}, \bibinfo {author} {\bibfnamefont
  {X.}~\bibnamefont {Zhang}}, \bibinfo {author} {\bibfnamefont
  {Z.}~\bibnamefont {Wang}}, \bibinfo {author} {\bibfnamefont {G.}~\bibnamefont
  {Cao}}, \bibinfo {author} {\bibfnamefont {T.}~\bibnamefont {Wang}}, \bibinfo
  {author} {\bibfnamefont {J.-J.}\ \bibnamefont {Zhang}}, \bibinfo {author}
  {\bibfnamefont {D.}~\bibnamefont {Culcer}}, \bibinfo {author} {\bibfnamefont
  {X.}~\bibnamefont {Hu}}, \bibinfo {author} {\bibfnamefont {H.-W.}\
  \bibnamefont {Jiang}}, \bibinfo {author} {\bibfnamefont {H.-O.}\ \bibnamefont
  {Li}}, \bibinfo {author} {\bibfnamefont {G.-C.}\ \bibnamefont {Guo}},\ and\
  \bibinfo {author} {\bibfnamefont {G.-P.}\ \bibnamefont {Guo}},\ }\bibfield
  {title} {\bibinfo {title} {Ultrafast coherent control of a hole spin qubit in
  a germanium quantum dot},\ }\href
  {https://doi.org/10.1038/s41467-021-27880-7} {\bibfield  {journal} {\bibinfo
  {journal} {Nature Communications}\ }\textbf {\bibinfo {volume} {13}},\
  \bibinfo {pages} {206} (\bibinfo {year} {2022})}\BibitemShut {NoStop}%
\bibitem [{\citenamefont {Kloeffel}\ \emph {et~al.}(2013)\citenamefont
  {Kloeffel}, \citenamefont {Trif}, \citenamefont {Stano},\ and\ \citenamefont
  {Loss}}]{PhysRevB.88.241405}%
  \BibitemOpen
  \bibfield  {author} {\bibinfo {author} {\bibfnamefont {C.}~\bibnamefont
  {Kloeffel}}, \bibinfo {author} {\bibfnamefont {M.}~\bibnamefont {Trif}},
  \bibinfo {author} {\bibfnamefont {P.}~\bibnamefont {Stano}},\ and\ \bibinfo
  {author} {\bibfnamefont {D.}~\bibnamefont {Loss}},\ }\bibfield  {title}
  {\bibinfo {title} {Circuit {QED} with hole-spin qubits in {G}e/{S}i nanowire
  quantum dots},\ }\href {https://doi.org/10.1103/PhysRevB.88.241405}
  {\bibfield  {journal} {\bibinfo  {journal} {Phys. Rev. B}\ }\textbf {\bibinfo
  {volume} {88}},\ \bibinfo {pages} {241405} (\bibinfo {year}
  {2013})}\BibitemShut {NoStop}%
\bibitem [{\citenamefont {Venitucci}\ and\ \citenamefont
  {Niquet}(2019)}]{PhysRevB.99.115317}%
  \BibitemOpen
  \bibfield  {author} {\bibinfo {author} {\bibfnamefont {B.}~\bibnamefont
  {Venitucci}}\ and\ \bibinfo {author} {\bibfnamefont {Y.-M.}\ \bibnamefont
  {Niquet}},\ }\bibfield  {title} {\bibinfo {title} {Simple model for
  electrical hole spin manipulation in semiconductor quantum dots: Impact of
  dot material and orientation},\ }\href
  {https://doi.org/10.1103/PhysRevB.99.115317} {\bibfield  {journal} {\bibinfo
  {journal} {Phys. Rev. B}\ }\textbf {\bibinfo {volume} {99}},\ \bibinfo
  {pages} {115317} (\bibinfo {year} {2019})}\BibitemShut {NoStop}%
\bibitem [{\citenamefont {Milivojevi\ifmmode~\acute{c}\else
  \'{c}\fi{}}(2021)}]{PhysRevB.104.235304}%
  \BibitemOpen
  \bibfield  {author} {\bibinfo {author} {\bibfnamefont {M.}~\bibnamefont
  {Milivojevi\ifmmode~\acute{c}\else \'{c}\fi{}}},\ }\bibfield  {title}
  {\bibinfo {title} {Electrical control of the hole spin qubit in {S}i and {G}e
  nanowire quantum dots},\ }\href {https://doi.org/10.1103/PhysRevB.104.235304}
  {\bibfield  {journal} {\bibinfo  {journal} {Phys. Rev. B}\ }\textbf {\bibinfo
  {volume} {104}},\ \bibinfo {pages} {235304} (\bibinfo {year}
  {2021})}\BibitemShut {NoStop}%
\bibitem [{\citenamefont {Adelsberger}\ \emph
  {et~al.}(2022{\natexlab{a}})\citenamefont {Adelsberger}, \citenamefont
  {Benito}, \citenamefont {Bosco}, \citenamefont {Klinovaja},\ and\
  \citenamefont {Loss}}]{PhysRevB.105.075308}%
  \BibitemOpen
  \bibfield  {author} {\bibinfo {author} {\bibfnamefont {C.}~\bibnamefont
  {Adelsberger}}, \bibinfo {author} {\bibfnamefont {M.}~\bibnamefont {Benito}},
  \bibinfo {author} {\bibfnamefont {S.}~\bibnamefont {Bosco}}, \bibinfo
  {author} {\bibfnamefont {J.}~\bibnamefont {Klinovaja}},\ and\ \bibinfo
  {author} {\bibfnamefont {D.}~\bibnamefont {Loss}},\ }\bibfield  {title}
  {\bibinfo {title} {Hole-spin qubits in {Ge} nanowire quantum dots: Interplay
  of orbital magnetic field, strain, and growth direction},\ }\href
  {https://doi.org/10.1103/PhysRevB.105.075308} {\bibfield  {journal} {\bibinfo
   {journal} {Phys. Rev. B}\ }\textbf {\bibinfo {volume} {105}},\ \bibinfo
  {pages} {075308} (\bibinfo {year} {2022}{\natexlab{a}})}\BibitemShut
  {NoStop}%
\bibitem [{\citenamefont {Adelsberger}\ \emph
  {et~al.}(2022{\natexlab{b}})\citenamefont {Adelsberger}, \citenamefont
  {Bosco}, \citenamefont {Klinovaja},\ and\ \citenamefont
  {Loss}}]{PhysRevB.106.235408}%
  \BibitemOpen
  \bibfield  {author} {\bibinfo {author} {\bibfnamefont {C.}~\bibnamefont
  {Adelsberger}}, \bibinfo {author} {\bibfnamefont {S.}~\bibnamefont {Bosco}},
  \bibinfo {author} {\bibfnamefont {J.}~\bibnamefont {Klinovaja}},\ and\
  \bibinfo {author} {\bibfnamefont {D.}~\bibnamefont {Loss}},\ }\bibfield
  {title} {\bibinfo {title} {Enhanced orbital magnetic field effects in {Ge}
  hole nanowires},\ }\href {https://doi.org/10.1103/PhysRevB.106.235408}
  {\bibfield  {journal} {\bibinfo  {journal} {Phys. Rev. B}\ }\textbf {\bibinfo
  {volume} {106}},\ \bibinfo {pages} {235408} (\bibinfo {year}
  {2022}{\natexlab{b}})}\BibitemShut {NoStop}%
\bibitem [{\citenamefont {Kloeffel}\ \emph {et~al.}(2011)\citenamefont
  {Kloeffel}, \citenamefont {Trif},\ and\ \citenamefont
  {Loss}}]{PhysRevB.84.195314}%
  \BibitemOpen
  \bibfield  {author} {\bibinfo {author} {\bibfnamefont {C.}~\bibnamefont
  {Kloeffel}}, \bibinfo {author} {\bibfnamefont {M.}~\bibnamefont {Trif}},\
  and\ \bibinfo {author} {\bibfnamefont {D.}~\bibnamefont {Loss}},\ }\bibfield
  {title} {\bibinfo {title} {Strong spin-orbit interaction and helical hole
  states in {Ge/Si} nanowires},\ }\href
  {https://doi.org/10.1103/PhysRevB.84.195314} {\bibfield  {journal} {\bibinfo
  {journal} {Phys. Rev. B}\ }\textbf {\bibinfo {volume} {84}},\ \bibinfo
  {pages} {195314} (\bibinfo {year} {2011})}\BibitemShut {NoStop}%
\bibitem [{\citenamefont {Li}(2021)}]{RL2021}%
  \BibitemOpen
  \bibfield  {author} {\bibinfo {author} {\bibfnamefont {R.}~\bibnamefont
  {Li}},\ }\bibfield  {title} {\bibinfo {title} {Low-energy subband
  wave-functions and effective g-factor of one-dimensional hole gas},\ }\href
  {https://doi.org/10.1088/1361-648x/ac0d18} {\bibfield  {journal} {\bibinfo
  {journal} {Journal of Physics: Condensed Matter}\ }\textbf {\bibinfo {volume}
  {33}},\ \bibinfo {pages} {355302} (\bibinfo {year} {2021})}\BibitemShut
  {NoStop}%
\bibitem [{\citenamefont {Li}(2022)}]{RL2022a}%
  \BibitemOpen
  \bibfield  {author} {\bibinfo {author} {\bibfnamefont {R.}~\bibnamefont
  {Li}},\ }\bibfield  {title} {\bibinfo {title} {Searching strong `spin'-orbit
  coupled one-dimensional hole gas in strong magnetic fields},\ }\href
  {https://doi.org/10.1088/1361-648x/ac37da} {\bibfield  {journal} {\bibinfo
  {journal} {Journal of Physics: Condensed Matter}\ }\textbf {\bibinfo {volume}
  {34}},\ \bibinfo {pages} {075301} (\bibinfo {year} {2022})}\BibitemShut
  {NoStop}%
\bibitem [{\citenamefont {Kloeffel}\ \emph {et~al.}(2018)\citenamefont
  {Kloeffel}, \citenamefont {Ran\ifmmode \check{c}\else
  \v{c}\fi{}i\ifmmode~\acute{c}\else \'{c}\fi{}},\ and\ \citenamefont
  {Loss}}]{PhysRevB.97.235422}%
  \BibitemOpen
  \bibfield  {author} {\bibinfo {author} {\bibfnamefont {C.}~\bibnamefont
  {Kloeffel}}, \bibinfo {author} {\bibfnamefont {M.~J.}\ \bibnamefont
  {Ran\ifmmode \check{c}\else \v{c}\fi{}i\ifmmode~\acute{c}\else \'{c}\fi{}}},\
  and\ \bibinfo {author} {\bibfnamefont {D.}~\bibnamefont {Loss}},\ }\bibfield
  {title} {\bibinfo {title} {Direct rashba spin-orbit interaction in {Si} and
  {Ge} nanowires with different growth directions},\ }\href
  {https://doi.org/10.1103/PhysRevB.97.235422} {\bibfield  {journal} {\bibinfo
  {journal} {Phys. Rev. B}\ }\textbf {\bibinfo {volume} {97}},\ \bibinfo
  {pages} {235422} (\bibinfo {year} {2018})}\BibitemShut {NoStop}%
\bibitem [{\citenamefont {Maier}\ \emph {et~al.}(2014)\citenamefont {Maier},
  \citenamefont {Klinovaja},\ and\ \citenamefont {Loss}}]{PhysRevB.90.195421}%
  \BibitemOpen
  \bibfield  {author} {\bibinfo {author} {\bibfnamefont {F.}~\bibnamefont
  {Maier}}, \bibinfo {author} {\bibfnamefont {J.}~\bibnamefont {Klinovaja}},\
  and\ \bibinfo {author} {\bibfnamefont {D.}~\bibnamefont {Loss}},\ }\bibfield
  {title} {\bibinfo {title} {Majorana fermions in {Ge/Si} hole nanowires},\
  }\href {https://doi.org/10.1103/PhysRevB.90.195421} {\bibfield  {journal}
  {\bibinfo  {journal} {Phys. Rev. B}\ }\textbf {\bibinfo {volume} {90}},\
  \bibinfo {pages} {195421} (\bibinfo {year} {2014})}\BibitemShut {NoStop}%
\bibitem [{\citenamefont {Luo}\ \emph {et~al.}(2017)\citenamefont {Luo},
  \citenamefont {Li},\ and\ \citenamefont {Zunger}}]{PhysRevLett.119.126401}%
  \BibitemOpen
  \bibfield  {author} {\bibinfo {author} {\bibfnamefont {J.-W.}\ \bibnamefont
  {Luo}}, \bibinfo {author} {\bibfnamefont {S.-S.}\ \bibnamefont {Li}},\ and\
  \bibinfo {author} {\bibfnamefont {A.}~\bibnamefont {Zunger}},\ }\bibfield
  {title} {\bibinfo {title} {Rapid transition of the hole rashba effect from
  strong field dependence to saturation in semiconductor nanowires},\ }\href
  {https://doi.org/10.1103/PhysRevLett.119.126401} {\bibfield  {journal}
  {\bibinfo  {journal} {Phys. Rev. Lett.}\ }\textbf {\bibinfo {volume} {119}},\
  \bibinfo {pages} {126401} (\bibinfo {year} {2017})}\BibitemShut {NoStop}%
\bibitem [{\citenamefont {Luo}\ \emph {et~al.}(2011)\citenamefont {Luo},
  \citenamefont {Zhang},\ and\ \citenamefont {Zunger}}]{PhysRevB.84.121303}%
  \BibitemOpen
  \bibfield  {author} {\bibinfo {author} {\bibfnamefont {J.-W.}\ \bibnamefont
  {Luo}}, \bibinfo {author} {\bibfnamefont {L.}~\bibnamefont {Zhang}},\ and\
  \bibinfo {author} {\bibfnamefont {A.}~\bibnamefont {Zunger}},\ }\bibfield
  {title} {\bibinfo {title} {Absence of intrinsic spin splitting in
  one-dimensional quantum wires of tetrahedral semiconductors},\ }\href
  {https://doi.org/10.1103/PhysRevB.84.121303} {\bibfield  {journal} {\bibinfo
  {journal} {Phys. Rev. B}\ }\textbf {\bibinfo {volume} {84}},\ \bibinfo
  {pages} {121303} (\bibinfo {year} {2011})}\BibitemShut {NoStop}%
\bibitem [{\citenamefont {Li}\ and\ \citenamefont {Zhang}(2022)}]{RL2022b}%
  \BibitemOpen
  \bibfield  {author} {\bibinfo {author} {\bibfnamefont {R.}~\bibnamefont
  {Li}}\ and\ \bibinfo {author} {\bibfnamefont {H.}~\bibnamefont {Zhang}},\
  }\bibfield  {title} {\bibinfo {title} {Electrical manipulation of a hole
  `spin'-orbit qubit in nanowire quantum dot: the nontrivial magnetic field
  effects},\ }\href
  {http://iopscience.iop.org/article/10.1088/1674-1056/ac873b} {\bibfield
  {journal} {\bibinfo  {journal} {Chinese Physics B}\ } (\bibinfo {year}
  {2022})}\BibitemShut {NoStop}%
\bibitem [{\citenamefont {Trif}\ \emph {et~al.}(2008)\citenamefont {Trif},
  \citenamefont {Golovach},\ and\ \citenamefont {Loss}}]{trif2008spin}%
  \BibitemOpen
  \bibfield  {author} {\bibinfo {author} {\bibfnamefont {M.}~\bibnamefont
  {Trif}}, \bibinfo {author} {\bibfnamefont {V.~N.}\ \bibnamefont {Golovach}},\
  and\ \bibinfo {author} {\bibfnamefont {D.}~\bibnamefont {Loss}},\ }\bibfield
  {title} {\bibinfo {title} {Spin dynamics in inas nanowire quantum dots
  coupled to a transmission line},\ }\href
  {https://doi.org/10.1103/PhysRevB.77.045434} {\bibfield  {journal} {\bibinfo
  {journal} {Phys. Rev. B}\ }\textbf {\bibinfo {volume} {77}},\ \bibinfo
  {pages} {045434} (\bibinfo {year} {2008})}\BibitemShut {NoStop}%
\bibitem [{\citenamefont {Li}\ \emph {et~al.}(2013)\citenamefont {Li},
  \citenamefont {You}, \citenamefont {Sun},\ and\ \citenamefont
  {Nori}}]{RL2013}%
  \BibitemOpen
  \bibfield  {author} {\bibinfo {author} {\bibfnamefont {R.}~\bibnamefont
  {Li}}, \bibinfo {author} {\bibfnamefont {J.~Q.}\ \bibnamefont {You}},
  \bibinfo {author} {\bibfnamefont {C.~P.}\ \bibnamefont {Sun}},\ and\ \bibinfo
  {author} {\bibfnamefont {F.}~\bibnamefont {Nori}},\ }\bibfield  {title}
  {\bibinfo {title} {Controlling a nanowire spin-orbit qubit via
  electric-dipole spin resonance},\ }\href
  {https://doi.org/10.1103/PhysRevLett.111.086805} {\bibfield  {journal}
  {\bibinfo  {journal} {Phys. Rev. Lett.}\ }\textbf {\bibinfo {volume} {111}},\
  \bibinfo {pages} {086805} (\bibinfo {year} {2013})}\BibitemShut {NoStop}%
\bibitem [{\citenamefont {Lutchyn}\ \emph {et~al.}(2010)\citenamefont
  {Lutchyn}, \citenamefont {Sau},\ and\ \citenamefont
  {Das~Sarma}}]{PhysRevLett.105.077001}%
  \BibitemOpen
  \bibfield  {author} {\bibinfo {author} {\bibfnamefont {R.~M.}\ \bibnamefont
  {Lutchyn}}, \bibinfo {author} {\bibfnamefont {J.~D.}\ \bibnamefont {Sau}},\
  and\ \bibinfo {author} {\bibfnamefont {S.}~\bibnamefont {Das~Sarma}},\
  }\bibfield  {title} {\bibinfo {title} {Majorana fermions and a topological
  phase transition in semiconductor-superconductor heterostructures},\ }\href
  {https://doi.org/10.1103/PhysRevLett.105.077001} {\bibfield  {journal}
  {\bibinfo  {journal} {Phys. Rev. Lett.}\ }\textbf {\bibinfo {volume} {105}},\
  \bibinfo {pages} {077001} (\bibinfo {year} {2010})}\BibitemShut {NoStop}%
\bibitem [{\citenamefont {Oreg}\ \emph {et~al.}(2010)\citenamefont {Oreg},
  \citenamefont {Refael},\ and\ \citenamefont {von
  Oppen}}]{PhysRevLett.105.177002}%
  \BibitemOpen
  \bibfield  {author} {\bibinfo {author} {\bibfnamefont {Y.}~\bibnamefont
  {Oreg}}, \bibinfo {author} {\bibfnamefont {G.}~\bibnamefont {Refael}},\ and\
  \bibinfo {author} {\bibfnamefont {F.}~\bibnamefont {von Oppen}},\ }\bibfield
  {title} {\bibinfo {title} {Helical liquids and majorana bound states in
  quantum wires},\ }\href {https://doi.org/10.1103/PhysRevLett.105.177002}
  {\bibfield  {journal} {\bibinfo  {journal} {Phys. Rev. Lett.}\ }\textbf
  {\bibinfo {volume} {105}},\ \bibinfo {pages} {177002} (\bibinfo {year}
  {2010})}\BibitemShut {NoStop}%
\bibitem [{\citenamefont {Lu}\ \emph {et~al.}(2005)\citenamefont {Lu},
  \citenamefont {Xiang}, \citenamefont {Timko}, \citenamefont {Wu},\ and\
  \citenamefont {Lieber}}]{Lu10046}%
  \BibitemOpen
  \bibfield  {author} {\bibinfo {author} {\bibfnamefont {W.}~\bibnamefont
  {Lu}}, \bibinfo {author} {\bibfnamefont {J.}~\bibnamefont {Xiang}}, \bibinfo
  {author} {\bibfnamefont {B.~P.}\ \bibnamefont {Timko}}, \bibinfo {author}
  {\bibfnamefont {Y.}~\bibnamefont {Wu}},\ and\ \bibinfo {author}
  {\bibfnamefont {C.~M.}\ \bibnamefont {Lieber}},\ }\bibfield  {title}
  {\bibinfo {title} {One-dimensional hole gas in germanium/silicon nanowire
  heterostructures},\ }\href {https://doi.org/10.1073/pnas.0504581102}
  {\bibfield  {journal} {\bibinfo  {journal} {Proceedings of the National
  Academy of Sciences}\ }\textbf {\bibinfo {volume} {102}},\ \bibinfo {pages}
  {10046} (\bibinfo {year} {2005})}\BibitemShut {NoStop}%
\bibitem [{\citenamefont {Gao}\ \emph {et~al.}(2020)\citenamefont {Gao},
  \citenamefont {Wang}, \citenamefont {Watzinger}, \citenamefont {Hu},
  \citenamefont {Ran{\v c}i{\'c}}, \citenamefont {Zhang}, \citenamefont {Wang},
  \citenamefont {Yao}, \citenamefont {Wang}, \citenamefont {Kuku{\v c}ka},
  \citenamefont {Vuku{\v s}i{\'c}}, \citenamefont {Kloeffel}, \citenamefont
  {Loss}, \citenamefont {Liu}, \citenamefont {Katsaros},\ and\ \citenamefont
  {Zhang}}]{Gao2020AM}%
  \BibitemOpen
  \bibfield  {author} {\bibinfo {author} {\bibfnamefont {F.}~\bibnamefont
  {Gao}}, \bibinfo {author} {\bibfnamefont {J.-H.}\ \bibnamefont {Wang}},
  \bibinfo {author} {\bibfnamefont {H.}~\bibnamefont {Watzinger}}, \bibinfo
  {author} {\bibfnamefont {H.}~\bibnamefont {Hu}}, \bibinfo {author}
  {\bibfnamefont {M.~J.}\ \bibnamefont {Ran{\v c}i{\'c}}}, \bibinfo {author}
  {\bibfnamefont {J.-Y.}\ \bibnamefont {Zhang}}, \bibinfo {author}
  {\bibfnamefont {T.}~\bibnamefont {Wang}}, \bibinfo {author} {\bibfnamefont
  {Y.}~\bibnamefont {Yao}}, \bibinfo {author} {\bibfnamefont {G.-L.}\
  \bibnamefont {Wang}}, \bibinfo {author} {\bibfnamefont {J.}~\bibnamefont
  {Kuku{\v c}ka}}, \bibinfo {author} {\bibfnamefont {L.}~\bibnamefont {Vuku{\v
  s}i{\'c}}}, \bibinfo {author} {\bibfnamefont {C.}~\bibnamefont {Kloeffel}},
  \bibinfo {author} {\bibfnamefont {D.}~\bibnamefont {Loss}}, \bibinfo {author}
  {\bibfnamefont {F.}~\bibnamefont {Liu}}, \bibinfo {author} {\bibfnamefont
  {G.}~\bibnamefont {Katsaros}},\ and\ \bibinfo {author} {\bibfnamefont
  {J.-J.}\ \bibnamefont {Zhang}},\ }\bibfield  {title} {\bibinfo {title}
  {Site-controlled uniform {Ge/Si} hut wires with electrically tunable
  spin--orbit coupling},\ }\href
  {https://doi.org/https://doi.org/10.1002/adma.201906523} {\bibfield
  {journal} {\bibinfo  {journal} {Advanced Materials}\ }\textbf {\bibinfo
  {volume} {32}},\ \bibinfo {pages} {1906523} (\bibinfo {year}
  {2020})}\BibitemShut {NoStop}%
\bibitem [{\citenamefont {Lawaetz}(1971)}]{PhysRevB.4.3460}%
  \BibitemOpen
  \bibfield  {author} {\bibinfo {author} {\bibfnamefont {P.}~\bibnamefont
  {Lawaetz}},\ }\bibfield  {title} {\bibinfo {title} {Valence-band parameters
  in cubic semiconductors},\ }\href {https://doi.org/10.1103/PhysRevB.4.3460}
  {\bibfield  {journal} {\bibinfo  {journal} {Phys. Rev. B}\ }\textbf {\bibinfo
  {volume} {4}},\ \bibinfo {pages} {3460} (\bibinfo {year} {1971})}\BibitemShut
  {NoStop}%
\bibitem [{\citenamefont {Csontos}\ \emph {et~al.}(2009)\citenamefont
  {Csontos}, \citenamefont {Brusheim}, \citenamefont {Z\"ulicke},\ and\
  \citenamefont {Xu}}]{PhysRevB.79.155323}%
  \BibitemOpen
  \bibfield  {author} {\bibinfo {author} {\bibfnamefont {D.}~\bibnamefont
  {Csontos}}, \bibinfo {author} {\bibfnamefont {P.}~\bibnamefont {Brusheim}},
  \bibinfo {author} {\bibfnamefont {U.}~\bibnamefont {Z\"ulicke}},\ and\
  \bibinfo {author} {\bibfnamefont {H.~Q.}\ \bibnamefont {Xu}},\ }\bibfield
  {title} {\bibinfo {title} {Spin-$\frac{3}{2}$ physics of semiconductor hole
  nanowires: Valence-band mixing and tunable interplay between bulk-material
  and orbital bound-state spin splittings},\ }\href
  {https://doi.org/10.1103/PhysRevB.79.155323} {\bibfield  {journal} {\bibinfo
  {journal} {Phys. Rev. B}\ }\textbf {\bibinfo {volume} {79}},\ \bibinfo
  {pages} {155323} (\bibinfo {year} {2009})}\BibitemShut {NoStop}%
\bibitem [{\citenamefont {Bychkov}\ and\ \citenamefont
  {Rashba}(1984)}]{bychkov1984oscillatory}%
  \BibitemOpen
  \bibfield  {author} {\bibinfo {author} {\bibfnamefont {Y.~A.}\ \bibnamefont
  {Bychkov}}\ and\ \bibinfo {author} {\bibfnamefont {E.~I.}\ \bibnamefont
  {Rashba}},\ }\bibfield  {title} {\bibinfo {title} {Oscillatory effects and
  the magnetic susceptibility of carriers in inversion layers},\ }\href
  {https://doi.org/10.1088/0022-3719/17/33/015} {\bibfield  {journal} {\bibinfo
   {journal} {Journal of Physics C: Solid State Physics}\ }\textbf {\bibinfo
  {volume} {17}},\ \bibinfo {pages} {6039} (\bibinfo {year}
  {1984})}\BibitemShut {NoStop}%
\bibitem [{\citenamefont {Nadj-Perge}\ \emph {et~al.}(2010)\citenamefont
  {Nadj-Perge}, \citenamefont {Frolov}, \citenamefont {Bakkers},\ and\
  \citenamefont {Kouwenhoven}}]{nadj2010spin}%
  \BibitemOpen
  \bibfield  {author} {\bibinfo {author} {\bibfnamefont {S.}~\bibnamefont
  {Nadj-Perge}}, \bibinfo {author} {\bibfnamefont {S.~M.}\ \bibnamefont
  {Frolov}}, \bibinfo {author} {\bibfnamefont {E.~P. A.~M.}\ \bibnamefont
  {Bakkers}},\ and\ \bibinfo {author} {\bibfnamefont {L.~P.}\ \bibnamefont
  {Kouwenhoven}},\ }\bibfield  {title} {\bibinfo {title} {Spin--orbit qubit in
  a semiconductor nanowire},\ }\href {https://doi.org/10.1038/nature09682}
  {\bibfield  {journal} {\bibinfo  {journal} {Nature}\ }\textbf {\bibinfo
  {volume} {468}},\ \bibinfo {pages} {1084} (\bibinfo {year}
  {2010})}\BibitemShut {NoStop}%
\bibitem [{\citenamefont {Li}(2018)}]{RL2018c}%
  \BibitemOpen
  \bibfield  {author} {\bibinfo {author} {\bibfnamefont {R.}~\bibnamefont
  {Li}},\ }\bibfield  {title} {\bibinfo {title} {A spin dephasing mechanism
  mediated by the interplay between the spin-orbit coupling and the
  asymmetrical confining potential in a semiconductor quantum dot},\ }\href
  {https://doi.org/10.1088/1361-648X/aadcb8} {\bibfield  {journal} {\bibinfo
  {journal} {Journal of Physics: Condensed Matter}\ }\textbf {\bibinfo {volume}
  {30}},\ \bibinfo {pages} {395304} (\bibinfo {year} {2018})}\BibitemShut
  {NoStop}%
\bibitem [{\citenamefont {Li}(2020)}]{RL2020}%
  \BibitemOpen
  \bibfield  {author} {\bibinfo {author} {\bibfnamefont {R.}~\bibnamefont
  {Li}},\ }\bibfield  {title} {\bibinfo {title} {Charge noise induced spin
  dephasing in a nanowire double quantum dot with spin{\textendash}orbit
  coupling},\ }\href {https://doi.org/10.1088/1361-648x/ab4933} {\bibfield
  {journal} {\bibinfo  {journal} {Journal of Physics: Condensed Matter}\
  }\textbf {\bibinfo {volume} {32}},\ \bibinfo {pages} {025305} (\bibinfo
  {year} {2020})}\BibitemShut {NoStop}%
\bibitem [{\citenamefont {Liao}\ \emph {et~al.}(2015)\citenamefont {Liao},
  \citenamefont {Luo}, \citenamefont {Yang}, \citenamefont {Chen},\ and\
  \citenamefont {Xu}}]{doi:10.1063/1.4929412}%
  \BibitemOpen
  \bibfield  {author} {\bibinfo {author} {\bibfnamefont {G.}~\bibnamefont
  {Liao}}, \bibinfo {author} {\bibfnamefont {N.}~\bibnamefont {Luo}}, \bibinfo
  {author} {\bibfnamefont {Z.}~\bibnamefont {Yang}}, \bibinfo {author}
  {\bibfnamefont {K.}~\bibnamefont {Chen}},\ and\ \bibinfo {author}
  {\bibfnamefont {H.~Q.}\ \bibnamefont {Xu}},\ }\bibfield  {title} {\bibinfo
  {title} {{Electronic structures of [001]- and [111]-oriented InSb and GaSb
  free-standing nanowires}},\ }\href {https://doi.org/10.1063/1.4929412}
  {\bibfield  {journal} {\bibinfo  {journal} {Journal of Applied Physics}\
  }\textbf {\bibinfo {volume} {118}},\ \bibinfo {pages} {094308} (\bibinfo
  {year} {2015})}\BibitemShut {NoStop}%
\end{thebibliography}%
\end{document}